\documentclass[conference]{IEEEtran}
\IEEEoverridecommandlockouts

\usepackage{amssymb,mathrsfs,amsmath}
\usepackage[linesnumbered,ruled,vlined]{algorithm2e}
\usepackage{graphicx}
\usepackage{algpseudocode}
\usepackage{lipsum}
\usepackage{setspace}
\usepackage{enumitem}
\usepackage{multirow}
\usepackage{makecell}
\usepackage{listings}
\usepackage{hyperref}[colorlinks, linkcolor=red, anchorcolor=blue, citecolor=green]
\usepackage[dvipsnames, svgnames, x11names]{xcolor}
\usepackage{pifont}
\usepackage{tcolorbox}
\usepackage{wasysym}
\usepackage{subfigure}
\usepackage{url}

\makeatletter
\let\old@lstKV@SwitchCases\lstKV@SwitchCases
\def\lstKV@SwitchCases#1#2#3{}
\makeatother
\usepackage{lstlinebgrd}
\makeatletter
\let\lstKV@SwitchCases\old@lstKV@SwitchCases

\lst@Key{numbers}{none}{%
    \def\lst@PlaceNumber{\lst@linebgrd}%
    \lstKV@SwitchCases{#1}%
    {none:\\%
     left:\def\lst@PlaceNumber{\llap{\normalfont
                \lst@numberstyle{\thelstnumber}\kern\lst@numbersep}\lst@linebgrd}\\%
     right:\def\lst@PlaceNumber{\rlap{\normalfont
                \kern\linewidth \kern\lst@numbersep
                \lst@numberstyle{\thelstnumber}}\lst@linebgrd}%
    }{\PackageError{Listings}{Numbers #1 unknown}\@ehc}}
\makeatother

\definecolor{verylightgray}{rgb}{.95,.95,.95}

\lstdefinelanguage{Solidity}{
	identifierstyle=\color{black},
	sensitive=false,
	comment=[l]{//},
	morecomment=[s]{/*}{*/},
	commentstyle=\color{gray}\footnotesize\ttfamily,
	stringstyle=\color{black}\ttfamily,
	morestring=[b]',
	morestring=[b]"
}

\lstdefinestyle{gitModify}
{
    language = Solidity,
	keywordstyle = {\color{red}}\bfseries,
	keywordstyle = [2]\color{blue}\bfseries,
	keywordstyle = [3]\color{teal}\bfseries,
	keywordstyle = [4]\color{violet}\bfseries,
	keywordstyle = [5]\color{Grey}\bfseries,
	otherkeywords = {++,--},
	morekeywords =[2]{anonymous, assembly, assert, break, call, callcode, case, catch, class, constant, continue, constructor, contract, debugger, default, delegatecall, delete, do, else, emit, event, experimental, export, external, false, finally, for, function, gas, if, implements, import, in, indexed, instanceof, interface, internal, is, length, library, log0, log1, log2, log3, log4, memory, modifier, new, payable, pragma, private, protected, public, pure, push, require, return, returns, revert, selfdestruct, send, solidity, storage, struct, suicide, super, switch, then, this, throw, true, try, typeof, using, value, view, while, with, addmod, ecrecover, keccak256, mulmod, ripemd160, sha256, sha3}, 
	morekeywords =[3]{address, bool, byte, bytes, bytes1, bytes2, bytes3, bytes4, bytes5, bytes6, bytes7, bytes8, bytes9, bytes10, bytes11, bytes12, bytes13, bytes14, bytes15, bytes16, bytes17, bytes18, bytes19, bytes20, bytes21, bytes22, bytes23, bytes24, bytes25, bytes26, bytes27, bytes28, bytes29, bytes30, bytes31, bytes32, enum, int, int8, int16, int24, int32, int40, int48, int56, int64, int72, int80, int88, int96, int104, int112, int120, int128, int136, int144, int152, int160, int168, int176, int184, int192, int200, int208, int216, int224, int232, int240, int248, int256, mapping, string, uint, uint8, uint16, uint24, uint32, uint40, uint48, uint56, uint64, uint72, uint80, uint88, uint96, uint104, uint112, uint120, uint128, uint136, uint144, uint152, uint160, uint168, uint176, uint184, uint192, uint200, uint208, uint216, uint224, uint232, uint240, uint248, uint256, var, void, ether, finney, szabo, wei, days, hours, minutes, seconds, weeks, years},	
	morekeywords =[4]{block, blockhash, coinbase, difficulty, gaslimit, number, timestamp, msg, data, gas, sender, sig, value, now, tx, gasprice, origin},	
	morekeywords =[5]{++,--},
}

\newcommand{\PP}[1]{
\vspace{2px}
\noindent{\bf \IfEndWith{#1}{.}{#1}{#1.}}
}

\begin{document}
\begin{sloppypar}

\title{MuFuzz: Sequence-Aware Mutation and Seed Mask Guidance for Blockchain Smart Contract Fuzzing}

\author{
        Peng Qian\IEEEauthorrefmark{2}\IEEEauthorrefmark{3}, 
	Hanjie Wu\IEEEauthorrefmark{4}, 
	Zeren Du\IEEEauthorrefmark{4}, 
	Turan Vural\IEEEauthorrefmark{5},  
	Dazhong Rong\IEEEauthorrefmark{2}, 
	Zheng Cao\IEEEauthorrefmark{2}\IEEEauthorrefmark{3}, \\
	Lun Zhang\IEEEauthorrefmark{3},
	Yanbin Wang\IEEEauthorrefmark{2}\IEEEauthorrefmark{3}, 
	Jianhai Chen\IEEEauthorrefmark{2}\IEEEauthorrefmark{1},
	Qinming He\IEEEauthorrefmark{2}\\
	\IEEEauthorrefmark{2}College of Computer Science and Technology, Zhejiang University, Hangzhou, China\\
	\IEEEauthorrefmark{3}Goplus Security, Hangzhou, China\\
	\IEEEauthorrefmark{4}School of Computer Science and Technology, Zhejiang Gongshang University, Hangzhou, China\\
 \IEEEauthorrefmark{5}University of California, Los Angeles, USA\\
       \thanks{\IEEEauthorrefmark{1} Jianhai Chen is the corresponding author.}
	pqian@zju.edu.cn, 18267186044@163.com, duzeren2021@gmail.com, turan@cs.ucla.edu\\
	\{rdz98, z.cao\}@zju.edu.cn, allen@gopluslabs.io, wybpaper@gmail.com, \{chenjh919, hqm\}@zju.edu.cn
}

\maketitle

\begin{abstract}
As blockchain smart contracts become more widespread and carry more valuable digital assets, they become an increasingly attractive target for attackers. Over the past few years, smart contracts have been subject to a plethora of devastating attacks, resulting in billions of dollars in financial losses. There has been a notable surge of research interest in identifying defects in smart contracts. However, existing smart contract fuzzing tools are still unsatisfactory. They struggle to screen out meaningful transaction sequences and specify critical inputs for each transaction. As a result, they can only trigger a limited range of contract states, making it difficult to unveil complicated vulnerabilities hidden in the deep state space.

In this paper, we shed light on smart contract fuzzing by employing a sequence-aware mutation and seed mask guidance strategy. In particular, we first utilize data-flow-based feedback to determine transaction orders in a meaningful way and further introduce a sequence-aware mutation technique to explore deeper states. Thereafter, we design a mask-guided seed mutation strategy that biases the generated transaction inputs to hit target branches. In addition, we develop a dynamic-adaptive energy adjustment paradigm that balances the fuzzing resource allocation during a fuzzing campaign. We implement our designs into a new smart contract fuzzer named MuFuzz, and extensively evaluate it on three benchmarks. Empirical results demonstrate that MuFuzz outperforms existing tools in terms of both branch coverage and bug finding. Overall, MuFuzz achieves higher branch coverage than state-of-the-art fuzzers (up to 25\%) and detects 30\% more bugs than existing bug detectors.
\end{abstract}


\section{Introduction}
Blockchain, a distributed and shared database, has garnered significant attention for its potential to revolutionize multiple industries and profoundly reshape the global landscape. Ethereum, one of the most widely-used blockchains, has been at the forefront of academic and industry attention as it endows a wide range of decentralized applications~\cite{qian2019digital,mirabelli2020blockchain,macrinici2018smart}. 

Ethereum enables the execution of so-called smart contracts, which are Turing-complete programs running on blockchains. Smart contracts have witnessed explosive adoption in recent years, with over 60 million instances created on Ethereum and over 1 million transactions completed per day~\cite{dune}. As smart contracts become more popular and manage more valuable digital assets, attackers are highly incentivized to find and exploit various vulnerabilities to steal profits. Consequently, bugs in smart contracts can lead to huge losses, as evidenced by recent attacks~\cite{Parity,atzei2017survey}. The recent spate of attack incidents on Ethereum smart contracts, e.g., FoMo3D~\cite{fomo3d}, Uniswap~\cite{uniswap}, and Cream Finance~\cite{Inspex}, has resulted in cumulative financial losses exceeding \$1 billion. To make matters worse, due to the immutability of the blockchain, it is extremely difficult to upgrade smart contracts, as old versions cannot be modified without manipulating 51\% of the blockchain's computing power to do so~\cite{lin2017survey}. In this context, it is crucial to vet the contract code pre-deployment and deploy a bug-free contract.

As countermeasures, a variety of tools have been developed to detect  smart contract bugs~\cite{qian2022smart}. Existing methods fall into two broad categories. One collection of works~\cite{luu2016making,tsankov2018securify,mythril} focus on static analysis (e.g., symbolic execution) to identify vulnerabilities. However, they either aggressively overestimate the execution of smart contracts and generate high false positives, or precisely enumerate symbolic traces of the entire contract and thus struggle to handle contracts that have many execution paths. Another group of efforts~\cite{he2019learning,jiang2018contractfuzzer,nguyen2020sfuzz} use dynamic analysis techniques (such as fuzzing) to uncover bugs in smart contracts. While fuzzing has shown promising progress in detecting program vulnerabilities, it suffers from the inherent difficulty of revealing bugs hidden in program regions not covered by the fuzzer, leading to false negatives. 

More specifically, when fuzzing smart contracts, we may face the following challenges. (1) Unlike traditional software, smart contracts are stateful programs that require a sequence of transactions as inputs to maintain the \emph{persistent} state, which refers to the data and variables permanently stored on the blockchain~\cite{choi2021smartian}. Current smart contract fuzzers still have difficulty in effectively identifying transaction sequences that can trigger the alteration to the \emph{persistent} state of a contract. Although some previous researches~\cite{choi2021smartian,torres2021confuzzius} have considered using data dependency analysis to determine transaction sequences, they struggle to deal with cases that require specific transaction sequences (a concrete example is presented in \S\ref{example}). (2) When fuzzing turns to input generation for transaction sequences, the input space can be significantly broad. However, existing smart contract fuzzers tend to arbitrarily mutate input bytes while ignoring critical parts of the input that should not be mutated, thus reducing the probability of hitting branches that require strict conditions to be found. (3) Finally, the energy allocated to each branch in fuzzing is usually unbalanced. Fuzzers may waste massive resources in fuzzing common branches while providing inadequate energy to deeply nested branches, making it difficult to dive into complex branches to explore deeper states.

To tackle the above challenges, we present a framework for smart contract fuzzing with the following components:
(1) First, we introduce a sequence-aware mutation strategy that determines a transaction sequence of a contract by analyzing the data flow dependencies of state variables. Furthermore, we enable the sequence mutation to take more care of critical transactions and extend the sequence, leading the fuzzer to dive into deep states (\S\ref{sequence_mutation}). 
(2) Second, we incorporate a mask-guided seed mutation strategy that approximates certain crucial parts of the inputs that should not be mutated, helping the fuzzer to evolve the generated inputs towards target branches (\S\ref{mask_mutation}). 
(3) We design a dynamic-adaptive energy adjustment mechanism to balance the fuzzing resources for each branch. In particular, we employ a lightweight abstract interpreter to analyze path prefixes and assign branch weights, allowing the fuzzer to dynamically allocate resources  based on the weight value of each branch during fuzzing (\S\ref{target_branch_revisting}).  

We design and implement MuFuzz, a novel fuzzing framework for smart contracts, which consists of a sequence-aware mutation module, a seed mask guidance module, and a dynamic energy adjustment module. We extensively evaluate MuFuzz on three different datasets (i.e., \ding{172} 21K real-world smart contracts from Ethereum, \ding{173} 155 unique vulnerable contracts collected from various sources, and \ding{174} 500 large contracts, each containing over 30,000 transactions in Ethereum), and demonstrate that our system outperforms existing tools in terms of both coverage and bug finding. Quantitatively, MuFuzz achieves up to 25\% higher branch coverage compared to state-of-the-art smart contract fuzzers and detects up to 30\% more vulnerabilities than other bug detectors.

In summary, we make the following key contributions:
\begin{itemize}[topsep=1pt, itemsep=1pt]
    \item We design a sequence-aware mutation and seed mask guidance strategy for blockchain smart contract fuzzing.
    \item We propose a new fuzzing framework for smart contracts that consists of three key components: a sequence-aware mutation, a mask-guided seed mutation, and a dynamic-adaptive energy adjustment, which increases the probability of exploring deep contract states. The proposed modules hold the potential to be transferable to the fuzzing of smart contracts on alternative blockchain platforms.
    \item We implement MuFuzz\footnote{Code is available at \url{https://github.com/Messi-Q/MuFuzz}} and conduct extensive experiments on three benchmarks. Not only does MuFuzz outperform state-of-the-art smart contract fuzzers in terms of both coverage and runtime, but it is also able to detect more vulnerabilities than existing bug detection tools.
    \item We release both our system and benchmarks to facilitate future research, in the spirit of open science.
\end{itemize}

\section{Background}
\label{background}

\renewcommand\arraystretch{1.0}
\begin{table*}
        \tiny
	\centering
	\caption{A summary of bug classes supported by each tool. Tools marked with \CIRCLE\ support detecting the bug, while those marked with \Circle\ do not. For each bug type: BD denotes block dependency ({e.g.,} block.timestamp, block.number); UD implies unprotected delegatecall; EF denotes Ether freezing; IO implies integer over-/under- flow; RE denotes reentrancy; US implies unprotected self-destruct; SE denotes strict Ether equality; TO implies transaction origin use; UE denotes unhandled exception.} 
	\vspace{-1.8em}
	\resizebox{0.999\textwidth}{!}{
		\begin{tabular}{ l l l c c c c c c c c c }
		\hline
		\multirow{2}{*}{\textbf{Tool}} & \multirow{2}{*}{\textbf{Type}} & \multirow{2}{*}{\textbf{Public Available}} & \multicolumn{9}{c}{\textbf{Vulnerability Type}}  \\
		\cline{4-12}
		~ & ~ & ~ & \textbf{BD} & \textbf{UD} & \textbf{EF} & \textbf{IO} & \textbf{RE} & \textbf{US} & \textbf{SE}& \textbf{TO} & \textbf{UE}  \\
		\hline
		 ContractFuzzer~\cite{jiang2018contractfuzzer}  &  Fuzzer  &  https://github.com/gongbell/ContractFuzzer  & \CIRCLE  &  \CIRCLE  & \CIRCLE  &  \Circle  &  \CIRCLE  &  \Circle &  \Circle  &   \Circle &  \CIRCLE   \\
		 ContraMaster~\cite{wang2019vultron,wang2020oracle}  &  Fuzzer  &  https://github.com/ntu-SRSLab/vultron  &  \Circle  &  \Circle  &  \Circle & \CIRCLE   & \CIRCLE   & \Circle  &  \Circle  &  \Circle  & \CIRCLE    \\
		 {Echidna}~\cite{grieco2020echidna}  &  Fuzzer  &  https://github.com/crytic/echidna  &  \Circle  &  \Circle  & \Circle  & \Circle   &  \Circle  & \Circle  &  \Circle  &   \Circle &  \CIRCLE   \\
		 {Reguard}~\cite{liu2018reguard}  &  Fuzzer  &  Not Available  &  \Circle  &  \Circle  & \Circle &  \Circle  &  \CIRCLE  &  \Circle &  \Circle  &   \Circle & \Circle    \\
		 {Harvey}~\cite{wustholz2020harvey,wustholz2020targeted}  &  Fuzzer  &  Not Available  & \Circle   &  \Circle  &  \Circle &  \CIRCLE  &  \CIRCLE  & \Circle  &  \Circle  &   \Circle &  \CIRCLE   \\
		 {sFuzz}~\cite{nguyen2020sfuzz}  &  Fuzzer   & https://github.com/duytai/sFuzz  &  \CIRCLE  & \CIRCLE   & \CIRCLE  &  \CIRCLE  &  \CIRCLE  & \Circle  &  \Circle  &   \Circle &  \CIRCLE   \\
		  {IR-Fuzz}~\cite{10018241}  &  Fuzzer  & https://github.com/Messi-Q/IR-Fuzz  &  \CIRCLE  & \CIRCLE   & \CIRCLE  &  \CIRCLE  &  \CIRCLE  & \Circle  &  \CIRCLE  &   \Circle &  \CIRCLE   \\
		 {Smartian}~\cite{choi2021smartian}  &  Fuzzer   &  https://github.com/SoftSec-KAIST/Smartian  & \CIRCLE   & \CIRCLE   & \CIRCLE  &  \CIRCLE  &  \CIRCLE  &  \CIRCLE & \Circle   &   \CIRCLE &  \CIRCLE   \\
		 ILF~\cite{he2019learning}  &  Fuzzer   & https://github.com/eth-sri/ilf  &  \CIRCLE  &  \CIRCLE  &  \CIRCLE &  \Circle  &  \Circle  & \CIRCLE  &  \Circle  &   \Circle & \CIRCLE    \\
		 {ConFuzzius}~\cite{torres2021confuzzius}  &   Fuzzer    &  https://github.com/christoftorres/ConFuzzius  &  \CIRCLE  &  \CIRCLE  & \CIRCLE  &  \CIRCLE  &  \CIRCLE  & \CIRCLE  & \Circle   &   \Circle &  \CIRCLE   \\
		 xFuzz~\cite{xue2022xfuzz}  &  Fuzzer    & https://github.com/ToolmanInside/xfuzz\_tool   & \Circle   &  \CIRCLE  &  \Circle &  \Circle  &  \CIRCLE  & \Circle  &  \Circle  &   \CIRCLE &  \Circle   \\
		 RLF~\cite{su2022effectively}  &  Fuzzer   & https://github.com/Demonhero0/rlf   & \CIRCLE   &  \CIRCLE  &  \CIRCLE &  \Circle  &  \Circle  & \CIRCLE  &  \Circle  &   \Circle &  \CIRCLE   \\
		 Oyente~\cite{luu2016making}  & Static  Analyzer  &  https://github.com/melonproject/oyente  &  \CIRCLE  &  \Circle  & \Circle  &  \CIRCLE  &  \CIRCLE  & \Circle  &  \Circle  &   \Circle &  \Circle   \\
		 Osiris~\cite{torres2018osiris} &  Static  Analyzer  &  https://github.com/christoftorres/Osiris  &  \CIRCLE  &  \Circle  & \Circle  &  \CIRCLE  & \CIRCLE   & \Circle  &  \Circle  &   \Circle & \Circle   \\
		 Mythril~\cite{mythril} &  Static  Analyzer  &  https://github.com/ConsenSys/mythril  &  \CIRCLE  &  \CIRCLE  & \Circle  & \CIRCLE   &  \CIRCLE  & \CIRCLE  &    \CIRCLE &   \CIRCLE &  \CIRCLE   \\
		 Slither~\cite{feist2019slither} &  Static  Analyzer  &  https://github.com/crytic/slither  & \CIRCLE   &  \CIRCLE  & \CIRCLE  &  \Circle  &  \CIRCLE  & \CIRCLE  & \CIRCLE &   \CIRCLE &  \CIRCLE   \\
		 Securify1.0~\cite{tsankov2018securify} &  Static  Analyzer  &  https://github.com/eth-sri/securify  &  \Circle  &  \Circle  & \Circle  &  \Circle  &  \CIRCLE  &   \Circle &  \Circle  &   \Circle &  \CIRCLE   \\
		 Manticore~\cite{mossberg2019manticore} &  Static  Analyzer  & https://github.com/trailofbits/manticore   &  \CIRCLE  &  \CIRCLE  & \Circle  &  \CIRCLE  & \CIRCLE   & \CIRCLE  & \Circle   &   \CIRCLE &   \CIRCLE  \\
		 Maian~\cite{nikolic2018finding} & Static  Analyzer   &  https://github.com/MAIAN-tool/MAIAN  &   \Circle &  \Circle  &  \CIRCLE &  \Circle  & \Circle   &  \CIRCLE &    \Circle &   \Circle & \Circle    \\
		 SmartCheck~\cite{tikhomirov2018smartcheck} & Static  Analyzer   &  https://github.com/smartdec/smartcheck &  \CIRCLE  &  \Circle  & \CIRCLE  &  \CIRCLE  &    \CIRCLE &  \Circle &  \Circle  &   \CIRCLE & \CIRCLE    \\
		 {Zeus}~\cite{kalra2018zeus} &  Static  Analyzer  &  Not Available  &  \CIRCLE  &  \Circle  &  \Circle  &  \CIRCLE  &  \CIRCLE  &  \Circle &  \Circle  &   \CIRCLE & \CIRCLE   \\
		 {VeriSmart}~\cite{so2020verismart} & Static  Analyzer   &  https://github.com/kupl/VeriSmart-public  & \Circle   &  \Circle  & \Circle  &  \CIRCLE  &  \Circle  & \Circle  &  \Circle  &   \Circle &  \Circle   \\
		 {Vandal}~\cite{brent2018vandal,grech2018madmax} & Static  Analyzer   &  https://github.com/usyd-blockchain/vandal  & \Circle   &  \Circle  & \Circle  &  \Circle  &  \CIRCLE  & \CIRCLE  & \Circle   &   \CIRCLE & \CIRCLE   \\
		 Sereum~\cite{rodler2018sereum} &  Static  Analyzer  &  Not Available  &  \Circle  &  \Circle  & \Circle  & \Circle   &  \CIRCLE  &  \Circle & \Circle   &  \Circle  & \Circle    \\
		 {teEther}~\cite{krupp2018teether} &  Static  Analyzer  &  https://github.com/nescio007/teether  &  \Circle  &  \CIRCLE  &  \Circle &  \Circle  &  \Circle  &  \CIRCLE &    \Circle & \Circle   & \Circle    \\
		 {Sailfish}~\cite{bose2022sailfish} & Static  Analyzer   &  https://github.com/ucsb-seclab/sailfish  &  \Circle  & \Circle   &  \Circle &  \Circle  &  \CIRCLE  & \Circle  &  \Circle  &  \Circle  &   \Circle  \\
		 DefectChecker~\cite{chen2021defectchecker} & Static  Analyzer   &  https://github.com/Jiachi-Chen/DefectChecker  &  \CIRCLE  &  \Circle  & \CIRCLE  & \Circle   & \CIRCLE   &  \Circle & \Circle   &   \CIRCLE & \CIRCLE    \\
		\hline
		\end{tabular} 
}
\label{tools_bugs}
\vspace{-3.0em}
\end{table*}

\subsection{Ethereum Smart Contract}
\label{ethereum_smart_contract}
The notion of smart contracts was originally put forth in 1994 by Nick Szabo~\cite{szabo1997formalizing}, who described the concept of a trustless system containing self-executing computer programs. However, the concept did not become a reality until the emergence of the Ethereum in 2014~\cite{wood2014ethereum}. Ethereum is currently the most popular blockchain, which employs the Ethereum Virtual Machine (EVM) to execute smart contracts. Smart contracts maintain {\small\ttfamily Storage}, which is organized as a key-value store to persistently hold state variables~\cite{choi2021smartian}. Since smart contracts are stored on the blockchain, they inherit certain properties. The immutability of the blockchain ensures that the dapp/program execution strictly adheres to the rules defined in the smart contract, which are unalterable once deployed on-chain. Attributed to the decentralized nature, smart contracts allow transactions to transpire between anonymous parties without relying on a trusted third party. To date, more than 60 million smart contracts have been created on Ethereum~\cite{dune}, giving rise to a variety of decentralized applications, such as decentralized finance (DeFi)~\cite{victor2021detecting}, non-fungible tokens (NFT)~\cite{wang2021non}, Internet of Things (IoT)~\cite{zhang2018smart}, healthcare~\cite{yu2019comparison}, crowdsale~\cite{zichichi2019likestarter}, and many others~\cite{qian2019digital,mirabelli2020blockchain,yu2021blockchain}.

\subsection{Smart Contract Vulnerability}
\label{smart_contract_vulnerability}
Smart contracts on the Ethereum blockchain have been subject to numerous destructive attacks. The most notable ones include the DAO attack in 2016~\cite{dao}, the Parity Multisig Wallet attack in 2017~\cite{Parity}, the Beauty Chain attack in 2018~\cite{bec}, the Cream Finance attack in 2021~\cite{Inspex}, and the Rari Fuse Pool attack in 2022~\cite{Rari}, which together resulted in huge financial losses. Prior works such as~\cite{chen2020defining,choi2021smartian} have studied and defined various defects in Ethereum smart contracts. Here, we mainly focus on detecting nine types of vulnerabilities, which are summarized in Table~\ref{tools_bugs}. The reasons for considering these bug classes are: (1) A large body of previous research has demonstrated that these vulnerabilities account for a significant portion of the existing bugs in the Ethereum ecosystem~\cite{luu2016making,tsankov2018securify}. (2) Over 90\% of the financial losses in Ethereum smart contracts are caused by these vulnerabilities~\cite{atzei2017survey,feist2019slither,chen2020survey}. (3) We have investigated the bug classes handled by existing smart contract security tools and found that these vulnerabilities are of the highest concern~\cite{choi2021smartian,he2019learning,torres2021confuzzius,tikhomirov2018smartcheck}. 

Following the guidelines of~\cite{choi2021smartian}, we consider merging closely related bug classes. For example, we merge {\small\ttfamily block timestamp dependency} and {\small\ttfamily block number dependency} into {\small\ttfamily block dependency}. We also consider {\small\ttfamily unchecked external call} and {\small\ttfamily gasless send} as subclasses of {\small\ttfamily unhandled exception}. We show how MuFuzz defines bug oracles to expose these vulnerabilities in \S\ref{bug_oracle}.

\subsection{Review of Existing Smart Contract Bug-Finding Tools}
\label{review_existing_tools}
We now study 27 smart contract bug finding tools, including fuzzers and static analyzers, which are summarized in Table~\ref{tools_bugs}. The third column indicates if a tool is publicly available, and lists the URL addresses of open-sourced projects. The remaining columns show bug classes supported by each tool.

While existing tools put a lot of effort in identifying various smart contract bugs, they still produce significant false positives or false negatives in practice. Static analyzers, {e.g.,} Oyente~\cite{luu2016making} and Securify~\cite{tsankov2018securify}, overestimate the execution of the contract and thus produce many false positives. Current smart contract fuzzers, e.g., ContraMaster~\cite{wang2020oracle} and {sFuzz}~\cite{nguyen2020sfuzz}, have difficulties in achieving high code coverage and thus are prone to false negatives. Other works use static analysis feedback to guide fuzzers and perform hybrid fuzzing on smart contracts. Compared to traditional fuzzers, hybrid fuzzers like {Smartian}~\cite{choi2021smartian} and {ConFuzzius}~\cite{torres2021confuzzius} are able to find relatively meaningful transaction sequences that can guide the fuzzer to cover complex branch conditions, leading to an improvement in code coverage. Inspired by prior efforts, we consider enhancing smart contact fuzzing by identifying a critical transaction sequence that is able to traverse more regions of the contract code. We develop a sequence-aware mutation and seed mask guidance strategy that allows the fuzzer to achieve higher coverage and mine deeper states.

\section{Overview}
\label{overview}
In this section, we begin by providing a motivating example that emphasizes the importance of generating a specific transaction sequence to find bugs. We then highlight the key challenges in generating such transaction sequences and discuss the limitations of existing fuzzers. Finally, we present our fuzzing policy and show how it handles the example.

\subsection{Motivating Example}
\label{example}
Considering the motivating example in Figure~\ref{code:example}, we present a simplified {\small\ttfamily Crowdsale} contract, which is a variant of an example in~\cite{he2019learning}. The contract has a {\small\ttfamily constructor} function, along with three normal functions, namely {\small\ttfamily invest}, {\small\ttfamily refund}, and {\small\ttfamily withdraw}. The {\small\ttfamily constructor} function runs once only when the deployer instantiates the contract, which sets the goal of the crowdsale to 100 ether (i.e., {\small\ttfamily goal} = 100 ether at line 10) and the invested money to 0 ({i.e.,} {\small\ttfamily invested} = 0 at line 11), and assigns the creator of the crowdsale as its owner ({i.e.,} {\small\ttfamily owner} = {\small\ttfamily msg.sender} at line 12). Users can invest in the {\small\ttfamily Crowdsale} contract by invoking the function {\small\ttfamily invest} when the crowdsale is active (i.e., {\small\ttfamily phase} = 0). The crowdsale ends when its goal is reached ({i.e.,} {\small\ttfamily invested} $\geq$ 100), in which case the owner of the contract is able to take out all funds by calling the function {\small\ttfamily withdraw}. Note that the {\small\ttfamily Crowdsale} contract allows users to refund their investment by calling the function {\small\ttfamily refund} when the crowdsale is in {\small\ttfamily phase} = 0.

\begin{figure}
	 \begin{center}
	 \linespread{0.72}
\begin{lstlisting}[language=Solidity,frame=lines,style=gitModify,linebackgroundcolor={
        }]
contract Crowdsale {
  // State Variables
  uint256 phase = 0;  // 0: Active, 1: Success
  uint256 goal;
  uint256 invested;
  address owner;
  mapping(address => uint256) invests;
   
  constructor() public {
    goal = 100 ether;
    invested = 0;
    owner = msg.sender;
  }
  function invest(uint256 donations) public payable {
    if (invested < goal) {
      invests[msg.sender] += donations;
      invested += donations;
      phase = 0;
    } else {
      phase = 1;
    }
  }
  function refund() public {
    if(phase == 0) {
      msg.sender.transfer(invests[msg.sender]);
      invests[msg.sender] = 0;
    }
  }
  function withdraw() public {
    if (phase == 1) {
      bug();  // There exists a bug
      owner.transfer(invested);
    }
  }
}
\end{lstlisting}
    \end{center}
       \vspace{-1.8em}
	\caption{A simplified Crowdsale contract derived from~\cite{he2019learning}.}
	\vspace{-1.9em}
	\label{code:example}
\end{figure}

\begin{figure*}
    \centering
    \includegraphics[width=17.9cm]{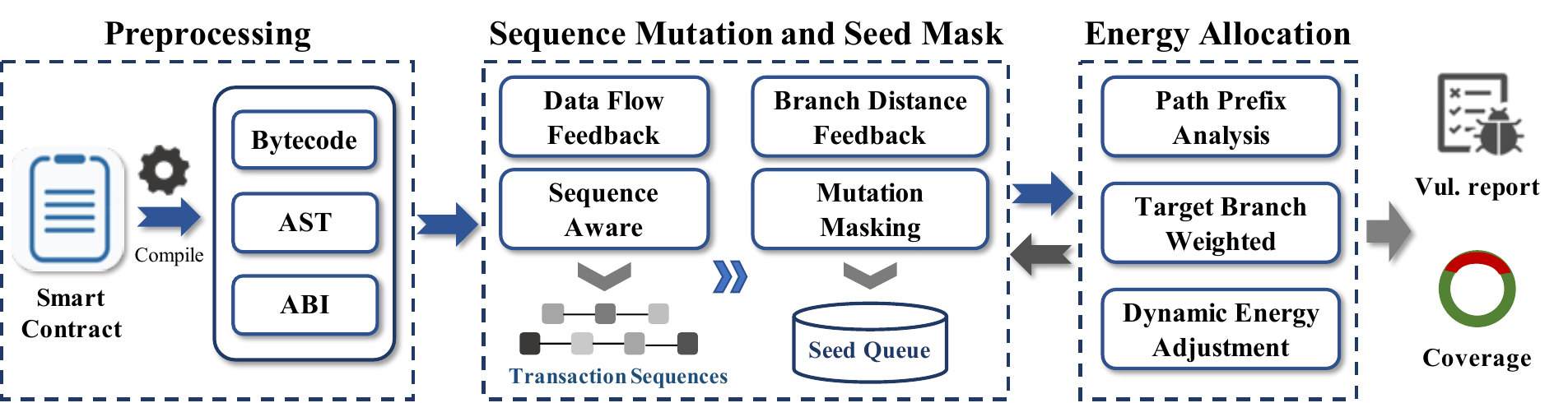}
     \vspace{-0.5em}
    \caption{A high-level architecture and analysis pipeline of {MuFuzz}. }
     \label{fig:overview}
     \vspace{-1.5em}
\end{figure*}

Unlike traditional programs, smart contracts take in a series of interconnected transactions as inputs to maintain the \emph{persistent} state~\cite{choi2021smartian}. As such, the unique challenge of fuzzing smart contracts is to find out a critical transaction sequence that can change the \emph{persistent} state. In particular, certain complex states within smart contracts can only be triggered by executing a specific sequence of transactions. For example, the function {\small\ttfamily withdraw} has a bug at line 31, which can be detected by a fuzzer only when the branch condition {\small\ttfamily phase} == 1 is satisfied. Achieving this goal is by no means trivial, however, as it requires the fuzzer not only to generate transactions in the correct order, but also to take into account the strict dependencies between transactions. For instance, assume that the fuzzer generates a transaction sequence [{\small\ttfamily withdraw} $\rightarrow$ {\small\ttfamily invest}($\ast$) $\rightarrow$ {\small\ttfamily refund}], where $\ast$ can be any value. Any attempt to mutate the test inputs of the functions will not trigger the bug because the state variable {\small\ttfamily phase} in the function {\small\ttfamily withdraw} has a read-dependency on the function {\small\ttfamily invest}. To find the bug, the fuzzer may need to generate a sequence such as [{\small\ttfamily invest}($\ast$) $\rightarrow$ {\small\ttfamily refund} $\rightarrow$ {\small\ttfamily withdraw}]. However, even with such a transaction sequence, the fuzzer fails to satisfy the branch condition at line 30, as calling the function {\small\ttfamily invest} once cannot enter the else-branch at line 20 to set {\small\ttfamily phase} = 1. To reach the else-branch, the function {\small\ttfamily invest} needs to be executed at least twice in the execution sequence, for example, [{\small\ttfamily invest}($\ast$) $\rightarrow$ {\small\ttfamily refund} $\rightarrow$ {\small\ttfamily invest}($\ast$) $\rightarrow$ {\small\ttfamily withdraw}]. That is, the first call achieves the crowdsale goal and the second call enters the else-branch at line 20. Next, we show the limitations of existing fuzzers for identifying such a specific sequence.

\subsection{Limitations of Existing Fuzzers}
\label{limitations}
We ran several state-of-the-art smart contract fuzzers on the example shown in Figure~\ref{code:example}, including {sFuzz}~\cite{nguyen2020sfuzz}, ILF~\cite{he2019learning}, {Smartian}~\cite{choi2021smartian}, and {ConFuzzius}~\cite{torres2021confuzzius}. None of these fuzzers were able to detect the bug of the contract at line 31, as they cannot navigate to the if-branch at line 30. In contrast, MuFuzz is able to tap into such a crucial branch and expose the bug within a matter of seconds.

After scrutinizing the implementations of existing smart contract fuzzers, we empirically found that they have inherent difficulties in determining critical transaction sequences, resulting in limited code coverage. Specifically, {sFuzz} generates transaction sequences by randomly mutating the order of transactions, making it difficult to identify a meaningful transaction sequence. ILF creates transaction sequences using a trained machine learning model. Since the machine learning algorithm is based on statistical reasoning, the generated sequence is not deterministic. While {Smartian} and {ConFuzzius} can potentially find transaction sequences in a relatively meaningful order by incorporating data dependency analysis, they are unable to handle the cases where a transaction must be executed consecutively, such as the example in Figure~\ref{code:example}. Moreover, current smart contract fuzzers tend to randomly mutate the test inputs of functions, resulting in the generation of seeds that have low probabilities of hitting a target branch. For example, {sFuzz} and {ConFuzzius} use crossover and mutation strategies from AFL~\cite{afl}, including byte flipping and interest value insertion, which apply a sequence of random mutations and thus are difficult to cover the complex branches guarded by strict conditions. In particular, they mutate bytes arbitrarily, ignoring which byte values are required to cover certain parts of the contract. Due to these challenges, existing fuzzers remain limited in the bugs they can find, as they struggle to explore deeper states in the contract.

\subsection{Our Fuzzing Scheme}
\label{fuzzing_policy}
To address the above limitations, we embrace three key designs in MuFuzz. 
\textbf{(1)} MuFuzz first collects a wide variety of relevant data flow dependencies to guide the generation of transaction sequences. MuFuzz figures out which state variables are defined in the contract, and analyzes the read and write dependencies of state variables between functions. Furthermore, we enable MuFuzz to determine whether a function has a read-after-write dependency on the same state variable \textbf{\textsc{v}} within itself and whether \textbf{\textsc{v}} is read by one of the branch statements. For example, satisfying the branch condition {\small\ttfamily invested} $<$ {\small\ttfamily goal} at line 15 of Figure~\ref{code:example} depends on the value of the state variable {\small\ttfamily invested} itself. Therefore, MuFuzz will perform a transaction sequence mutation to enforce functions that contain such variables to be executed consecutively, allowing the fuzzer to explore more complex states. 
\textbf{(2)} MuFuzz employs a branch-distance-feedback seed selection algorithm used by {sFuzz}~\cite{nguyen2020sfuzz}, which elects test inputs as seeds according to how far a test input is from satisfying the condition of a target branch. Preferably, to facilitate the seed evolution to approach the deeply nested branches, we incorporate a mask-guided seed mutation strategy into MuFuzz, which approximates certain parts of the inputs that should not be mutated. Then, MuFuzz does not allow mutating these crucial parts of the inputs, thus increasing the probability of hitting deep and complex branches. 
\textbf{(3)} MuFuzz employs a dynamic energy adjustment mechanism to guide the fuzzer to bias fuzzing energy to critical target branches, such as deeply nested branches and branches that may contain vulnerabilities.

\section{Design of MuFuzz}
\label{design}
We now present the design details of MuFuzz, with its high-level architecture shown in Figure~\ref{fig:overview}. 
MuFuzz consists of three key components: a sequence-aware mutation, a mask-guided seed mutation, and a dynamic-adaptive energy adjustment. In the following, we will describe each component in detail.

\begin{figure}
    \centering
    \includegraphics[width=8.4cm]{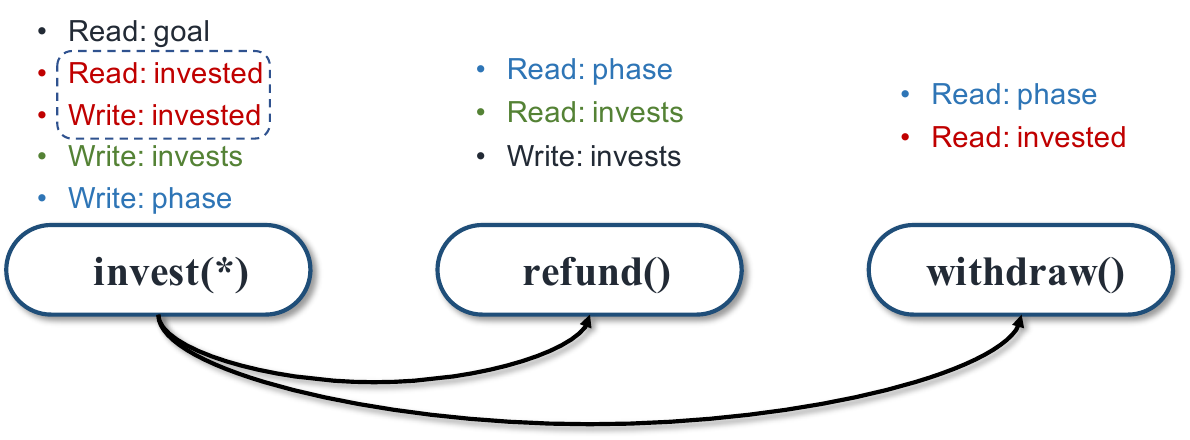}
    \vspace{-0.75em}
    \caption{A simple dependency graph to show the read and write dependencies included in the example of Figure~\ref{code:example}. }
     \label{fig:data-flow}
     \vspace{-1.8em}
\end{figure}

 \vspace{-0.3em}
\subsection{Sequence-Aware Mutation}
\label{sequence_mutation}
MuFuzz begins by taking the contract source code as inputs, which is then compiled into three types of representations, i.e., bytecode, application binary interface (ABI), and abstract syntax tree (AST). Bytecode is disassembled into EVM instructions for fuzzing. Meanwhile, MuFuzz captures the data dependencies of all state variables in the contract. By analyzing the ABI and AST, MuFuzz is able to figure out which state variables are defined and which functions contain state variables. Since smart contracts are stateful programs, MuFuzz ignores functions that do not contain any state variables because they cannot affect the \emph{persistent} state. MuFuzz then tracks each state variable and its read and write operations, such as assignments and comparisons. 

Afterwards, MuFuzz derives a transaction sequence based on the information gathered from the data dependency analysis of the state variables. Put succinctly, MuFuzz approximately determines a sequence of transactions in which transaction $T_1$ is executed before transaction $T_2$ only if $T_1$ writes a state variable \textbf{\textsc{v}} where $T_2$ reads it. As a result, MuFuzz is able to estimate the invocation order of each transaction in the sequence.
For example, we depict a simple dependency graph in Figure~\ref{fig:data-flow} to show the read and write dependencies of the contract {\small\ttfamily Crowdsale} in Figure~\ref{code:example}. MuFuzz learns that {\small\ttfamily invest($\ast$)} writes the state variables {\small\ttfamily invested}, {\small\ttfamily invests}, {\small\ttfamily phase}, while {\small\ttfamily refund} reads {\small\ttfamily phase}, {\small\ttfamily invests}, and {\small\ttfamily withdraw} reads {\small\ttfamily phase}, {\small\ttfamily invested}. Therefore, MuFuzz generates a transaction sequence like $\mathcal{S}$: [{\small\ttfamily invest}($\ast$) $\rightarrow$ {\small\ttfamily refund} $\rightarrow$ {\small\ttfamily withdraw}]. Note that MuFuzz places the {\small\ttfamily constructor} function at the beginning of the transaction sequence by default.

However, such a sequence still fails to satisfy the branch condition at line 30 of Figure~\ref{code:example}, because calling the function {\small\ttfamily invest} once cannot enter the else-branch at line 20 to set {\small\ttfamily phase} = 1. To address this challenge, we further consider a sequence mutation that enforces a function to be executed repeatedly in the sequence only if there is a read-after-write (RAW) dependency on a state variable \textbf{\textsc{v}} that is inside the function, and \textbf{\textsc{v}} is read by one of the branch statements. The dependency graph in Figure~\ref{fig:data-flow} shows the RAW dependencies of the state variable {\small\ttfamily invested} in the function {\small\ttfamily invest($\ast$)}. We observe that {\small\ttfamily invested} in the if-branch of the function {\small\ttfamily invest} (line 15) depends on itself at line 17, i.e., {\small\ttfamily invested} += {\small\ttfamily donations}. Therefore, MuFuzz will mutate the original sequence $\mathcal{S}$ to $\mathcal{S}_m$: [{\small\ttfamily invest}($\ast$) $\rightarrow$ {\small\ttfamily refund} $\rightarrow$ {\small\ttfamily invest}($\ast$) $\rightarrow$ {\small\ttfamily withdraw}]. Note that MuFuzz does not take into account local variables or function argument variables, as they impose no impact on the \emph{persistent} state. With the mutated sequence, MuFuzz may generate transaction inputs as [{\small\ttfamily invest}(100) $\rightarrow$ {\small\ttfamily refund} $\rightarrow$ {\small\ttfamily invest}(50) $\rightarrow$ {\small\ttfamily withdraw}]. Since the goal of the crowdsale is 100, the first call to {\small\ttfamily invest}(100) in $\mathcal{S}_m$ satisfies the condition of the if-branch at line 15, and the second call to {\small\ttfamily invest}(50) enters the else-branch at line 20 to set {\small\ttfamily phase} = 1. As a result, when {\small\ttfamily withdraw} is executed in $\mathcal{S}_m$, the fuzzer can enter the if-branch at line 30 and expose the potential bug. With the sequence-aware mutation strategy, MuFuzz effectively infers a meaningful transaction sequence, expanding the scope of explored branches and states.

\subsection{Mask-Guided Seed Mutation}
\label{mask_mutation}
We now determine the inputs of the transaction sequence. A trivial way is to randomly generate the test inputs of transactions. However, due to the randomness, it suffers from inherent difficulties in satisfying complicated branch conditions. Consider the example shown in Figure~\ref{code:example2}. By randomly mutating the test inputs, the probability of satisfying the branch condition ({i.e.,} {\small\ttfamily msg.value} == {\small\ttfamily 88 finney} at line 6) is as low as $\frac{1}{2^{256}}$ (assuming that each value is equally generated). Moreover, it is extremely difficult for existing smart contract fuzzers to enter deeply nested branches such as the if-branch at line 9 of Figure~\ref{code:example2}. To address these challenges, we introduce a seed evolution paradigm that iteratively refines the test inputs of transactions. The detailed procedures are summarized in Algorithm~\ref{alg1}. We first adopt a branch-distance-feedback seed selection strategy, guiding the fuzzer to select high-quality seeds. Furthermore, we design a mask-guided seed mutation strategy, which allows the fuzzer to identify the certain parts of the test inputs that should not be mutated, thus guiding the seed mutation to hit target branches more efficiently. MuFuzz starts by creating an empty seed queue and a set of seeds as inputs (lines 1-2), followed by performing seed selection (lines 4-13) and seed mutation (lines 14-29), respectively.

\begin{figure}
	 \begin{center}
	 \linespread{0.75}
	 \vspace{-0.25em}
\begin{lstlisting}[language=Solidity,frame=lines,style=gitModify,linebackgroundcolor={
        }]
contract Game {
  mapping(address => uint256) balance;
  
  function guessNum(uint256 number) public payable {
    uint256 random = uint256(keccak256(abi.encodePacked(block.timestamp, now))) % 200;
    require(msg.value == 88 finney);
    if (number < random) {
      uint256 luckyNum = number % 2;
      if (luckyNum == 0) {
        // There may exist an integer overflow bug
        balance[msg.sender] += msg.value * 10;
      } else {
        balance[msg.sender] += msg.value * 5;
      }
    } 
  }
}
\end{lstlisting}
    \end{center}
    \vspace{-1.8em}
	\caption{A simplified game contract of guess number.}
	\vspace{-2.0em}
	\label{code:example2}
\end{figure}

\textbf{Branch Distance Feedback.}\quad MuFuzz tracks the seed execution and records the branches that each test case covers. Whenever a test case covers a new branch, it is added to the seed queue. While this strategy has been shown to quickly traverse most of the branches, it still has difficulty in covering those branches that are guarded by strict conditions. For example, the probability of satisfying the condition {\small\ttfamily msg.value} == {\small\ttfamily 88 finney} at line 6 of Figure~\ref{code:example2} is tremendously low. Intuitively, however, it is obvious that a test input with {\small\ttfamily msg.value} == {\small\ttfamily 100 finney} is closer to meeting the condition than a test input with {\small\ttfamily msg.value} == {\small\ttfamily 1,000 finney}. Towards this, we integrate a branch-distance-feedback strategy into MuFuzz, which makes the fuzzer select seeds according to the measure of how far a test input is from covering any uncovered branch. Namely, the closer a test input $t$ is to covering the branch $b_r$, the smaller the distance is. With such feedback, MuFuzz is able to quickly approach those hard-to-cover deep branches.

\begin{algorithm}[t]
    \footnotesize    
    \setstretch{0.75} 
    \caption{ {\small \textsc{Seed Selection and Mutation} }}
    \label{alg1}
    
    \textsc{seedQueue} $\leftarrow \varnothing$\;
    $Seeds \leftarrow$ \textsc{initSeed}()\;
    
    \While{$\lnot${\rm\textsc{terminated()}}}{
       \ForEach{$seed \in Seeds$}{
          \If{$seed$ covers a new branch}{
              \textsc{seedQueue}.\textsc{add}($seed$)\;
          }
       }
       {\tcp*[h]{\emph{Branch-Distance-Feedback Seed Selection}} } \\
        
        Let $BR_{\textit{uncover}}$ be uncovered branches\;   
        \ForEach{$b_r \in BR_{\textit{uncover}}$}{ 
        Let $min$ be +$\infty$\;  
            \ForEach{$seed \in$ $Seeds$ }{
                \If{{\rm \textsc{distance}(}$seed, b_r${\rm )} $< min$}{
                 Let $min$ be \textsc{distance}($seed, b_r$)\; 
                }
            }
            \textsc{seedQueue}.\textsc{add}($seed$)\;
        }
       {\tcp*[h]{\emph{Mask-Guided Seed Mutation}} } \\
       $energy \leftarrow$ \textsc{assignMutationEnergy}()\;
        \While{$energy >$ {\rm0}}{
            \ForEach{$seed \in$ {\rm \textsc{seedQueue}}}{
                 \If{{\rm(}$seed$ does not cover a nested branch{\rm)} {\rm or (\textsc{distance(}}$seed$, $b_r${\rm)} does not decrease{\rm)}}{ 
                      continue\;
                 }         
                 $nestedBranch \leftarrow \textsc{nestedHit}(seed)$\;
                 $mask \leftarrow \textsc{computeMask}(seed, nestedBranch, BR_{uncover})$\;
            
            \ForEach{{\rm 0} $\leq$ i $\leq$ {\rm $|seed|$}}{
                 \ForEach{$mutation$ $\in$ $mutationTypes$}{
                   \If{$\lnot${\rm\textsc{okToMutate($mask, mutation, i$)}}}{
                         continue\;
                    }
                    $newseed$ $\leftarrow$ \textsc{mutate}($seed, mutation, i$)\;
                    $result$ $\leftarrow$ \textsc{fuzzRun}($newseed$)\;
                    \If{{\rm\textsc{newCoverage(}}$result${\rm)} } {
                        \textsc{seedQueue}.\textsc{add}($newseed$)\;
                    }
                    $energy$ $\leftarrow$ {\rm\textsc{updateEnergy}}($result$, $energy$)\;
               }
          }
           }
        }
    }
     \Return{{\rm \textsc{seedQueue}}}\;
\end{algorithm}

\textbf{Mutation Masking.}\quad To further bias test input generation towards target branches, we incorporate a mask-guided seed mutation into MuFuzz. The mutation masking strategy derives from the observations that: (1) certain parts of a test input that hits a deeply nested branch are critical to satisfying the necessary conditions for reaching that branch; (2) certain parts of a test input that makes the distance to cover a branch smaller play a key role in approaching that branch. Therefore, to generate more mutated inputs hitting target branches, the crucial parts of the test inputs should not be mutated. Inspired by this, MuFuzz first customizes the selection of test inputs to mutate from the seed queue. It selects the inputs that either hit the deeply nested branches or make the branch distance smaller. We say that a branch $b_r$ is a nested branch if and only if $b_r$ contains at least two nested conditional statements. Each nested branch is associated with a nested score \textit{nested\_score}, which is set to the number of nested conditional statements.

After filtering out which seeds need to be mutated, MuFuzz introduces a \emph{mutation mask} computation algorithm inspired by FariFuzz~\cite{lemieux2018fairfuzz}, aiming to approximate the critical parts of the test inputs that are not allowed to mutate. MuFuzz engages a set of mutation operators, including byte flipping, replacing bytes with interesting values, byte insertion, and byte deletion. Note that MuFuzz internally represents each test input $t$ as a byte stream. {A mutation operation is a tuple $m = (x, n)$, where $n$ represents the number of bytes affected by the mutation and $x$ denotes one of the following mutation types:}
\begin{itemize}[wide=10pt, topsep=1.5pt, itemsep=1pt, leftmargin=\dimexpr\labelwidth + 8.5\labelsep\relax]
    \item[\textbf{O:}] overwriting $n$ bytes with some values at position $i$ of $t$.
    \item[\textbf{I :}] inserting $n$ bytes of values at position $i$ of $t$.
    \item[\textbf{R:}] replacing $n$ bytes with some values at position $i$ of $t$.
    \item[\textbf{D:}] deleting $n$ bytes of values at position $i$ of $t$.
\end{itemize}

To select specific mutation types with $x \in \{\textbf{O}, \textbf{I}, \textbf{R}, \textbf{D}\}$, the values that are operated over existing bytes must be determined. For a test input $t$, a mutation $m = (x, n)$, and a position $i \in [0, |t| - n]$, we say that $\text{\textsc{MUTATE}}(t, m, i)$ indicates that a mutated input is generated by applying a mutation operation $m$ to a test input $t$ at position $i$. Algorithm~\ref{alg2} outlines how to compute a \emph{mutation mask} for a given test input $t$ and a corresponding target branch $b_r$. Specifically, for each mutation type $x$ that is executed on a test input $t$, MuFuzz determines whether a mutated input hits a nested branch or reduces the distance to reach the uncovered branch (lines 7, 10, 13, 16). If the conditions are satisfied, MuFuzz marks the position $i$ as overwritable (\textbf{O}), insertable (\textbf{I}), replaceable (\textbf{R}), and deletable (\textbf{D}), respectively (lines 8, 11, 14, 17). In the mutation stage (lines 21-29 in Algorithm~\ref{alg1}), MuFuzz first screens out mutants that could violate the test target via \textsc{okToMutate} (line 23 in Algorithm~\ref{alg1}). Then, MuFuzz adds a mutated input to the seed queue only if it contributes to an improvement in coverage. It is worth mentioning that the process of seed mutation continues until the mutation energy upper-bound is reached. Here, energy is defined as the number of times seeds are mutated at this stage. Next, we further illustrate how MuFuzz flexibly allocates fuzzing resources during a fuzzing campaign.

\begin{algorithm}[t]
  \setstretch{0.75}
  \footnotesize
    {
    \caption{ {\small \textsc{ComputeMask}(\textit{seed}, \textit{branch}, \textit{$BR_{uncover}$}) }}
    \label{alg2}
    \textit{mask} $\leftarrow$ \textsc{initEmptyMask}($|seed|$)\;
     \textit{n} $\leftarrow$ \textsc{Random}(1, $|seed|$)\;
     \textit{x} $\in$ \{\textbf{O}, \textbf{I}, \textbf{R}, \textbf{D}\}\;
    \ForEach{{\rm 0} $\leq$ i $\leq$ {\rm $|seed|$}}{
            \textit{m} $\leftarrow$  (\textit{x}, \textit{n})\;
           \textit{seed\_\textbf{O}} $\leftarrow$ \textsc{mutate}(\textit{seed, m, i})\;
           \If{\textit{branch} $\in$ {\rm\textsc{nestedHit}(}$seed\_\textbf{O}${\rm )}\ {\rm or \textsc{distance}(}$seed\_\textbf{O}$, \textit{$BR_{\textit{uncover}}$}{\rm )}\ decrease}{
               \textit{mask}[$i$] $\leftarrow$ \textit{mask}[$i$] $\cup$ \{\textbf{O}\}\;
           }
           \textit{seed\_\textbf{I}} $\leftarrow$ \textsc{mutate}(\textit{seed, m, i})\;
            \If{\textit{branch} $\in$ {\rm\textsc{nestedHit}(}seed\_\textbf{I}{\rm )}\ {\rm or \textsc{distance}(}seed\_\textbf{I}, \textit{$BR_{\textit{uncover}}$}{\rm )}\ decrease}{
               \textit{mask}[$i$] $\leftarrow$ \textit{mask}[$i$] $\cup$ \{\textbf{I}\}\;
           }
           \textit{seed\_\textbf{R}} $\leftarrow$ \textsc{mutate}(\textit{seed, m, i})\;
          \If{\textit{branch} $\in$ {\rm\textsc{nestedHit}(}seed\_\textbf{R}{\rm )}\ {\rm or \textsc{distance}(}seed\_\textbf{R}, \textit{$BR_{\textit{uncover}}$}{\rm )}\ decrease}{
               \textit{mask}[$i$] $\leftarrow$ \textit{mask}[$i$] $\cup$ \{\textbf{R}\}\;
           }
           \textit{seed\_\textbf{D}} $\leftarrow$ \textsc{mutate}(\textit{seed, m, i})\;
           \If{\textit{branch} $\in$ {\rm\textsc{nestedHit}(}seed\_\textbf{D}{\rm )}\ {\rm or \textsc{distance}(}seed\_\textbf{D}, \textit{$BR_{\textit{uncover}}$}{\rm )}\ decrease}{
               \textit{mask}[$i$] $\leftarrow$ \textit{mask}[$i$] $\cup$ \{\textbf{D}\}\;
           }
    }
   \Return{$mask$}\;
   }
\end{algorithm}

\subsection{Fuzzing with Dynamic Energy Adjustment}
\label{target_branch_revisting}
In practice, after reviewing a large number of real-world smart contracts, we empirically observe that the updating of state variables tends to be protected by strict branch conditions or hidden in deeply nested branches, such as {\small\ttfamily phase} at line 20 of Figure~\ref{code:example} and {\small\ttfamily balance[msg.sender]} at lines 11, 13 of Figure~\ref{code:example2}. Moreover, these complex branches may also harbor potential vulnerabilities, such as the possible integer overflow bug at line 11 of Figure~\ref{code:example2}. Unfortunately, conventional fuzzers may waste massive resources in fuzzing common branches, while the allocated energy is insufficient for the deeply nested branches or branches that are likely to contain bugs. To address this problem, we design a dynamic-adaptive energy adjustment mechanism, {which enables the fuzzing resource allocation for each branch more balanced and flexible.}

MuFuzz is equipped with a pre-fuzz phase that executes a test input on an instrumented EVM to collect the exercised path $\pi$ (line 1 in Algorithm~\ref{alg3}). Given the path $\pi$, MuFuzz initializes the fuzzing resources. After that, it analyzes all split points ({i.e.,} branch instruction {\small\ttfamily JUMPI}) in $\pi$. The loop from lines 6 to 16 respectively computes the number of nested conditional statements for each branch and checks if there is a branch that covers a vulnerable instruction (which we refer to instructions that may introduce vulnerabilities, e.g., {\small\ttfamily call.value}, {\small\ttfamily block.timestamp}). Here, we adopt a path prefix analysis technique~\cite{wustholz2020targeted} to help determine whether all vulnerable instruction locations (i.e., \emph{instLoc}) are reachable (lines 12-13). During the pre-fuzz phase, MuFuzz will set a weight value for each exercised branch. Note that the nested branches are assigned different weight values based on the value of \textit{nested\_score}, and the branch covering a vulnerable instruction is assigned an additional weight value. It is worth mentioning that the pre-fuzz phase yields little impact on the overall runtime overhead of the fuzzer.

\begin{algorithm}[t]
  \setstretch{0.75}
  \footnotesize
    \caption{ {\small \textsc{BranchWeighted}(\textit{seed}, \textit{instLoc}) }}
    \label{alg3}

     $\pi \leftarrow \textsc{preFuzzRun}(seed)$; \ {\scriptsize{\tcp*[h]{\emph{Pre-Fuzz for Collecting Path}}} } \\
     $\pi_{\textit{resource}} \leftarrow \textsc{initResource}(\pi)$; \ {\scriptsize{\tcp*[h]{\emph{Init Fuzzing Resource}}} } \\
     $(\textsc{B}_{\textit{nested}}, \mathcal{W}_1, \mathcal{N}_s) \leftarrow (\varnothing, \varnothing, \varnothing)$\;
     $(\textsc{B}_{\textit{vulnerable}}, \mathcal{W}_2) \leftarrow (\varnothing, \varnothing)$\;
     $\textit{nested\_score} \leftarrow 0 $\;
    \ForEach{$i < |\pi|$}{
    \If{{\rm \textsc{isBranchInstruction}}{\rm(}$i, \pi${\rm)}}{
    $\textit{nested\_score}\leftarrow\textit{nested\_score}+1$\;
    $w_1 \leftarrow \textsc{weightAssign}(\textit{nested\_score})$\;
    $(\textsc{B}_{\textit{nested}}, \mathcal{W}_1, \mathcal{N}_s).\textsc{add}(\pi [i], w_1, \textit{nested\_score})$\;
    $\pi_{\textit{pre}} \leftarrow \pi [0...i + 1]$\;
    $\phi, \textit{loc} \leftarrow \textsc{prefixInference}(\pi_{\textit{pre}})$; {\scriptsize{\tcp*[h]{\emph{Prefix Analysis}}} } \\
    \If{{\rm \textsc{isVulnerableInstructReached}}$(\phi, \textit{loc}, \textit{instLoc})$}{
        $w_2 \leftarrow \textsc{weightAssign}()$\;
        ($\textsc{B}_{\textit{vulnerable}}, \mathcal{W}_2).\textsc{add}(\pi [i], w_2)$\;
    }
    }
    $i \leftarrow i+1$\;
    }
   \Return{  ($\textsc{B}_{\textit{nested}}$, $\mathcal{W}_1, \mathcal{N}_s$), $(\textsc{B}_{\textit{vulnerable}}, \mathcal{W}_2)$  }\;
\end{algorithm}

In subsequent fuzzing rounds, MuFuzz dynamically adjusts resource allocation according to the weight value of each branch. This suggests that the higher the weight value of a target branch, the more fuzzing resources will be allocated along the path to that branch. Moreover, MuFuzz also leverages the energy allocation feedback to guide seed mutation, namely the seeds that reach branches covering the vulnerable instructions are preferentially selected and fuzzed. With the assistance of the dynamic-adaptive energy allocation strategy, MuFuzz is able to take care of these target branches, making the fuzzing process more balanced for each branch.

\subsection{Bug Oracle Implementation}
\label{bug_oracle}
Now, we briefly discuss the implementation of bug oracles for the nine classes of bugs defined in MuFuzz.

\textbf{Block Dependency (BD).}\quad First, we detect the block dependency by checking whether the execution trace contains an instruction that captures the block state, such as {\small\ttfamily TIMESTAMP}, {\small\ttfamily NUMBER}. Then, we monitor if such instructions contaminate {\small\ttfamily CALL}, {\small\ttfamily JUMPI}, or compare instructions ({e.g.,} {\small\ttfamily LT}, {\small\ttfamily GT}, {\small\ttfamily EQ}).

\textbf{Unprotected Delegatecall (UD).}\quad We detect the unprotected delegatecall by first checking if the execution trace contains a {\small\ttfamily DELEGATECALL} instruction. Then, we verify whether a modifier constraint exerts on the function that contains the {delegatecall} statement. Finally, we check if the argument of delegatecall is included in the function.

\textbf{Integer Over-/Under- Flow (IO).}\quad When an arithmetic operation produces a value that is outside the range of the integer type, an integer over-/under- flow vulnerability occurs. In such a scenario, the Ethereum Virtual Machine (EVM) will truncate the overflow bits. We detect the bug class by first checking if the execution trace contains an {\small\ttfamily ADD}, {\small\ttfamily MUL}, or {\small\ttfamily SUB} instruction. Then, we verify whether the result of the arithmetic operation is truncated in the EVM stack.

\textbf{Reentrancy (RE).}\quad Reentrancy can be simply considered as a contract invoking an external contract that can call back the original contract. As such, we  first check if the execution trace contains a {\small\ttfamily CALL} instruction whose gas value can be larger than 2,300 units ({i.e.,} a \emph{call.value} invocation). Then, we expose the reentrancy vulnerability by monitoring whether the function containing the \emph{call.value} invocation is called repeatedly during a fuzzing campaign.

\textbf{Strict Ether Equality (SE).}\quad When Ether balance is treated as the branch condition, a strict Ether equality vulnerability may occur. We detect the bug class by first checking whether the execution trace contains a {\small\ttfamily BALANCE} instruction. Then, we monitor whether the {\small\ttfamily BALANCE} instruction is followed by the conditional jump instruction {\small\ttfamily JUMPI} or compare instructions ({e.g.,} {\small\ttfamily LT}, {\small\ttfamily GT}, {\small\ttfamily EQ}).

\textbf{Unhandled Exception (UE).}\quad We identify the unhandled exception by first checking if the execution trace contains a call or a chain of calls ({i.e.,} {\small\ttfamily CALL} instruction). We then check whether the exception instruction ({i.e.,} {\small\ttfamily INVALID} instruction) occurs in every invocation of the call chain. Finally, we confirm if the result of the call chain flows into a conditional jump instruction {\small\ttfamily JUMPI}. 

For the rest of the bug classes, we implement the same bug oracles as {Smartian} (e.g., \textbf{TO}), {ContractFuzzer} (e.g., \textbf{EF}), and {ConFuzzius} (e.g., \textbf{US}), respectively.

\section{Evaluation}
\label{evaluation}
In this section, we carry out a series of experiments to evaluate the effectiveness and practicality of MuFuzz. We attempt to answer the following research questions. 
\begin{itemize}[wide=0pt, topsep=1pt, itemsep=1pt, leftmargin=\dimexpr\labelwidth + 5.5\labelsep\relax]
    \item[\textbf{RQ1.}] Does MuFuzz achieve higher branch coverage compared to existing state-of-the-art smart contract fuzzers?
    \item[\textbf{RQ2.}] Can MuFuzz effectively identify different types of vulnerabilities in smart contracts? How is its bug-finding performance against state-of-the-art detection tools?
    \item[\textbf{RQ3.}] How much do the individual components of MuFuzz contribute to its performance gains in terms of both branch coverage and vulnerability detection?
    \item[\textbf{RQ4.}] How does MuFuzz perform on real-world smart contracts involving large transactions in Ethereum? 
\end{itemize}

In what follows, we first introduce the experimental setup, followed by addressing the above questions one by one. In addition, we provide a case study to show the practicality of MuFuzz in detecting bugs in real-world smart contracts.

\subsection{Experimental Setup}
\label{experimental_setup}
\textbf{Datasets.} We conduct our experiments on three different datasets, which are summarized in Table~\ref{datasets}. The first dataset \textbf{D1} is used to measure the branch coverage achieved by the investigated fuzzers. The second dataset \textbf{D2} aims to evaluate the performance of the studied tools in detecting vulnerabilities, while the third dataset \textbf{D3} is specifically used to validate the capabilities of our system in handling real-world contracts with large-scale transactions. In the spirit of open science, we make all these datasets publicly available.
\begin{itemize}[wide=0pt, topsep=1pt, itemsep=1pt, leftmargin=\dimexpr\labelwidth + 4 \labelsep\relax]
    \item[\textbf{D1.}] First, we use the dataset published by~\cite{torres2021confuzzius}, which consists of 21,147 real-world Ethereum contracts. In particular, to measure how each tool performs as the size of the contract increases, this dataset is divided into two sub-datasets based on the number of compiled contract bytecode instructions—\textbf{small} ($\leq$ 3,632 instructions) and \textbf{large} ($>$ 3,632 instructions) datasets, which contain 17,803 and 3,344 contracts, respectively.
    \item[\textbf{D2.}] We also construct a benchmark dataset that covers nine types of vulnerabilities, which are listed in Table~\ref{tools_bugs}. We build this dataset by collecting vulnerable contracts from various sources, including SmartBugs~\cite{durieux2020empirical}, {VeriSmart}~\cite{so2020verismart}, TMP~\cite{zhuangsmart}, and SWC registry~\cite{swcRegistry}. We also collect contracts from blog posts that analyze contract vulnerabilities. As a result, we obtain a total of 155 unique contracts containing 217 annotated vulnerabilities. Note that a single contract may have multiple bug classes. 
    \item[\textbf{D3.}] This dataset consists of 500 popular and complex smart contracts published by~\cite{choi2021smartian}. All of these contracts are obtained from Etherscan~\cite{etherscan}, which is a decentralized platform that provides statistics on Ethereum smart contracts. Each contract contains more than 30,000 transactions.
\end{itemize}

\renewcommand\arraystretch{1.35}
\begin{table}
\small
\centering
\caption{Illustration of three benchmark datasets.}
\vspace{-0.9em}
\resizebox{0.49\textwidth}{!}{
\begin{tabular}{llll}
   \hline
   \textbf{\#} & \textbf{Source} & \textbf{Used For} & \textbf{Numbers of Contracts} \\
   \hline
   \textbf{D1} & {ConFuzzius}~\cite{torres2021confuzzius} & \textbf{RQ1, RQ3} & 17,803 small + 3,344 large  \\
   \textbf{D2} & \makecell[l]{{VeriSmart}~\cite{so2020verismart}, TMP~\cite{zhuangsmart} \\ SmartBugs~\cite{durieux2020empirical}, SWC registry~\cite{swcRegistry}} & \textbf{RQ2} & 155 vulnerable \\
   \textbf{D3} & {Smartian}~\cite{choi2021smartian} & \textbf{RQ4} & 500 popular  \\
    \hline
\end{tabular}
}
\label{datasets}
\vspace{-1.9em}
\end{table}

\begin{figure*}
\centering
\subfigure[Branch coverage of different methods on small contracts]{
\includegraphics[width=8.1cm]{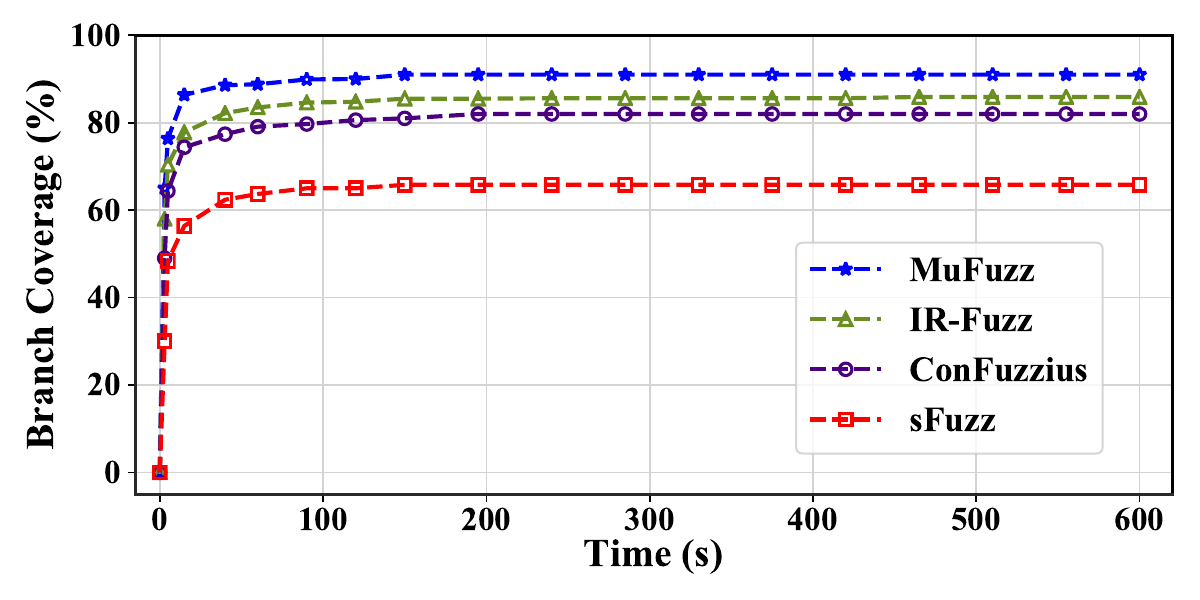}
} \quad\quad\;
\subfigure[Branch coverage of different methods  on large contracts]{
\includegraphics[width=8.1cm]{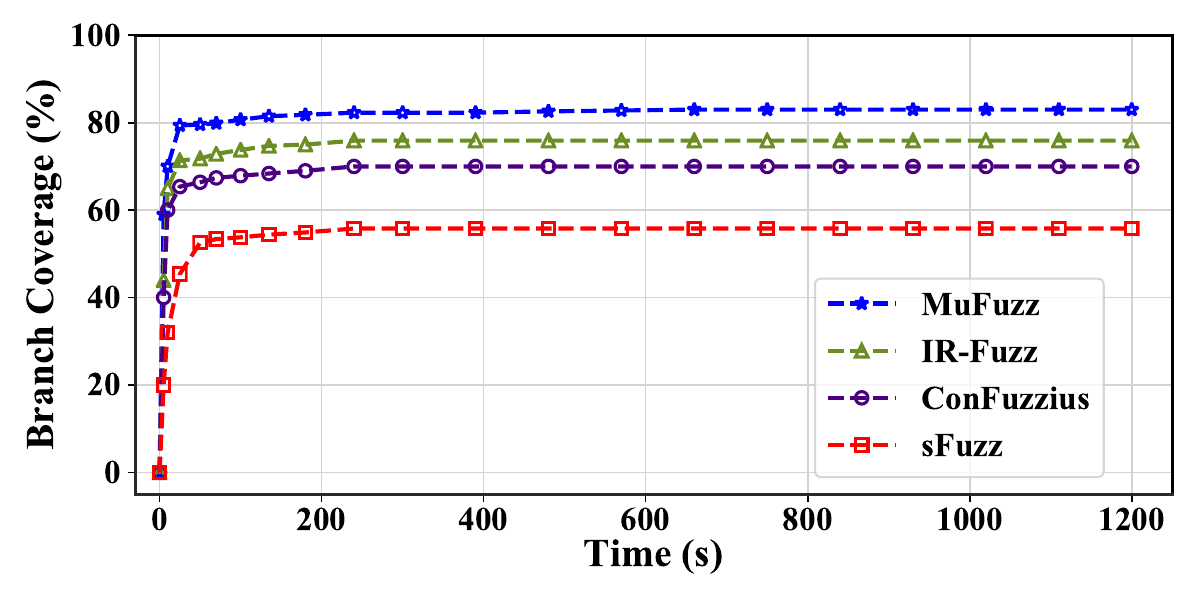}
}
\vspace{-0.75em}
\caption{Branch coverage achieved by MuFuzz and other fuzzers over time on small contracts and large contracts, respectively. }
\label{coverage_time}
\vspace{-1.75em}
\end{figure*}

In the experiments, we follow the setup of previous work such as~\cite{10018241,torres2021confuzzius} to determine the fuzzing time budgets. For \textbf{D1}, we conduct experiments on small contracts for 10 minutes each, and on large contracts for 20 minutes each. We run experiments on \textbf{D2} for 10 minutes per contract, and on \textbf{D3} for 20 minutes per contract.

\textbf{Comparing Baselines.} 
For static analyzers, we choose Oyente~\cite{luu2016making}, Mythril~\cite{mythril}, Osiris~\cite{torres2018osiris}, Securify~\cite{tsankov2018securify}, and Slither~\cite{feist2019slither}. 
Oyente is one of the pioneering bug detectors for smart contracts and enjoys popularity in the community. Mythril and Slither are commonly used by the industry and have been shown in a recent study~\cite{durieux2020empirical} that they outperform a variety of existing tools ({e.g.,} Manticore~\cite{mossberg2019manticore}, Maian~\cite{nikolic2018finding}, SmartCheck~\cite{tikhomirov2018smartcheck}, etc.). Osiris is one of the derivatives of Oyente, and Securify is one of the earliest tools to adopt the custom intermediate representation of Datalog for smart contract analysis. We compare MuFuzz to these static analyzers in terms of bug finding performance. {Note that static analysis tools do not actually execute smart contracts, and thus they do not provide code coverage information.}

For fuzzing tools, we have included {sFuzz}~\cite{lemieux2018fairfuzz}, {ConFuzzius}~\cite{torres2021confuzzius}, {Smartian}~\cite{choi2021smartian}, and IR-Fuzz~\cite{10018241} because: 1) their experiments have demonstrated superior performance than existing smart contract fuzzers, such as {ContractFuzzer}~\cite{jiang2018contractfuzzer}, {Echidna}~\cite{grieco2020echidna}, {ContraMaster}~\cite{wang2020oracle}, and ILF~\cite{he2019learning}, and 2) these tools are commonly regarded as comparing baselines by existing research. We compare MuFuzz to these fuzzers in terms of branch coverage, efficiency, and vulnerability detection performance. We ignore the comparisons with several fuzzers that are either not open source we ignore the comparisons with current smart contract fuzzers that are either not open source (e.g., {Harvey}~\cite{wustholz2020harvey}, {Reguard}~\cite{liu2018reguard}) or support detecting only a small set of bug classes (e.g., {xFuzz}~\cite{xue2022xfuzz}). Notably, we also make use of SmartBugs~\cite{durieux2020empirical} to perform comparative experiments, which is a framework that integrates existing representative analyzers for Ethereum smart contracts.

\textbf{Analysis Setup.} We run the experiments of MuFuzz on a machine running Ubuntu 18.04.6 LTS server, with two Inter(R) Core(TM) E5-2630 v3 CPUs at 2.40GHz (32 cores) and 256GB of memory. To compile contracts, we use the Solidity version of {\small\ttfamily solc-0.4.26}.

\renewcommand\arraystretch{1.0}
\begin{table*}
\centering
\small
\caption{The number of true positives / false negatives / timeout or error cases raised by each tool. False negatives represent that a tool fails to detect the corresponding bugs, and `n/a' indicates that a tool does not support detecting the bug class.}
   \vspace{-1.1em}
    \resizebox{0.99\textwidth}{!}{
    \begin{tabular}{c|ccccc|cccccc}
   \hline
    \multirow{2}*{\textbf{Types}}  &  \multicolumn{5}{c|}{\textbf{Static Analyzers}} &  \multicolumn{5}{c}{\textbf{Fuzzers}} \\
    \cline{2-6} \cline{7-11}
    ~ & \textbf{Oyente}  & \textbf{Mythril}  &  \textbf{Osiris}   &  \textbf{Securify} &  \textbf{Slither} &\textbf{{sFuzz}}  & \textbf{{ConFuzzius}}  & \textbf{{Smartian}}  &  \textbf{IR-Fuzz} &  \textbf{MuFuzz}  \\
       \hline
      \textbf{BD} &     5 / 14 / 1    &      8 / 5 / 7            &     8 / 11 / 1            &   n/a      &   3 / 16 / 1   &   10 / 10 / 0  &   10 / 4 / 6   &  11 / 9 / 0   &  13 / 7 / 0   & \textbf{15 / 5 / 0}   \\
      \textbf{UD} &    n/a                &        5 / 6 / 6          &       n/a          &   n/a          &   10 / 7 / 0    &     10 / 7 / 0       &  10 / 7 / 0         &  2 / 15 / 0    &  17 / 0 / 0  & \textbf{17 / 0 / 0}   \\
      \textbf{EF} &       n/a             & n/a                              & n/a                & n/a     &    7 / 15 / 0   &  7 / 15 / 0   &  6 / 14 / 2        & 0 / 22 / 0     & 13 / 9 / 0   &  \textbf{14 / 8 / 0}  \\
      \textbf{IO}  &   55 / 8 / 2    & 33 / 12 / 20                 &   44 / 20 / 1  & n/a              &  n/a   &  39 / 26 / 0       &   45 / 10 / 10                 & 43 / 22 / 0    & 50 / 15 / 0  & \textbf{62 / 3 / 0}  \\
      \textbf{RE} &   8 / 8 / 0     & 10 / 0 / 6                & 10 / 6 / 0    &  9 / 7 / 0   & 11 / 5 / 0   &  10 / 6 / 0      & 11 / 3 / 2                   & 4 / 12 / 0   & 12 / 4 / 0   &  \textbf{16 / 0 / 0} \\
      \textbf{US} &   n/a  & 10 / 8 / 5                & n/a                 & n/a           &  13 / 10 / 0     & n/a                 & 12 / 9 / 2    & 13 / 10 / 0   & n/a   & \textbf{23 / 0 / 0}  \\
      \textbf{SE} &       n/a             & 0 / 9 / 10                & n/a                 & n/a         &  3 / 16 / 0         & n/a                & n/a                   & n/a                  &  11 / 8 / 0                 &  \textbf{19 / 0 / 0}  \\
      \textbf{TO} &   n/a                & 2 / 0 / 0                 & n/a                 & n/a          & 2 / 0 / 0        & n/a                & n/a                   & 2 / 0 / 0                   & n/a  & \textbf{2 / 0 / 0}  \\
      \textbf{UE} &   n/a     & 10 / 3 / 18                 &   n/a                  & 17 / 14 / 0  & 2 / 29 / 0 &    12 / 19 / 0    & 16 / 13 / 2        & 19 / 12 / 0   &  20 / 11 / 0 &  \textbf{27 / 4 / 0}  \\
      \hline
      \textbf{Total} &   68 / 30 / 3  & 78 / 43 / 72 & 62 / 37 / 2 & 26 / 21 / 0  & 51 / 98 / 1 & 88 / 83 / 0 & 110 / 60 / 24  &  94 / 102 / 0 &  136 / 54 / 0 & \textbf{195 / 20 / 0}   \\
     \hline
      \end{tabular}
      }
      \vspace{-2.0em}
    \label{results}
\end{table*}

 \vspace{-0.3em}
\subsection{Branch Coverage (Answering RQ1)}
\label{code_coverage}
We begin by measuring the code coverage achieved by MuFuzz and other fuzzers. The main metric we measure is the number of basic block transitions covered, which is also referred to as branch coverage. We conduct the coverage experiments on \textbf{D1-small} and \textbf{D1-large}, respectively. Specifically, we count the number of different branches covered by generated test inputs from the seed queue. 

\begin{figure}
    \centering
    \includegraphics[width=8.1cm]{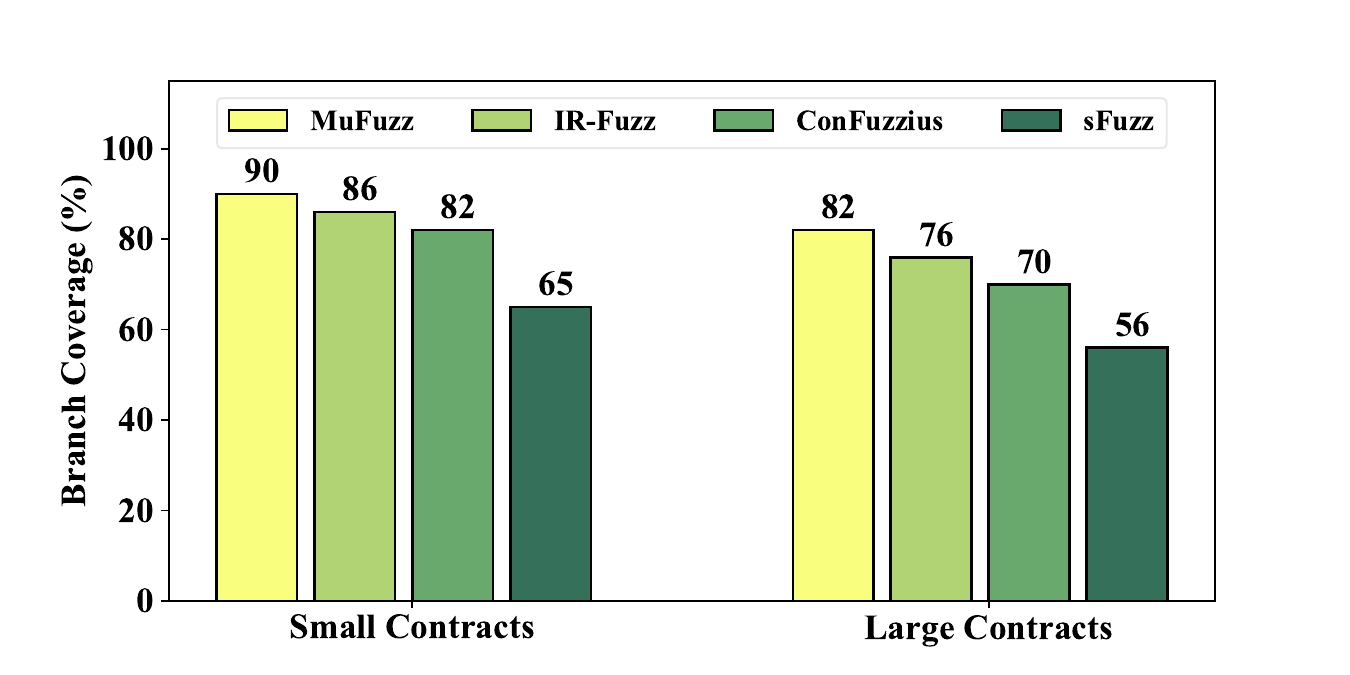}
    \vspace{-1.0em}
    \caption{Overall coverage of MuFuzz and other fuzzers. }
     \label{coverage_overview}
     \vspace{-2.0em}
\end{figure}

We compare MuFuzz with sFuzz, ConFuzzius, and IR-Fuzz, where Figure~\ref{coverage_overview} shows the average branch coverage on the clusters of small and large contracts individually. Note that we ignore the coverage comparison with Smartian as branch coverage data is not available in its fuzzing report. It is evident that MuFuzz achieves the highest branch coverage on both small and large contracts, with 90\% and 82\%, respectively. As expected, each tool achieves relatively low coverage on large contracts compared to small contracts. However, while the other fuzzers tested exhibited a significant coverage slippage, MuFuzz's slippage was only approximately 8\%. Further, we plot the tendency of the branch coverage achieved by each fuzzer over time in Figure~\ref{coverage_time}. From the figure, we can observe that not only does MuFuzz consistently outperform the other fuzzers, but it also achieves higher coverage in a shorter time. On average, MuFuzz achieves 81\% branch coverage on small contracts after 10 seconds, which is 29\%, 12\%, and 7\% higher than {sFuzz}, {ConFuzzius}, and IR-Fuzz, respectively. On the large contracts, MuFuzz produces 70\% branch coverage after 10 seconds, while the other three fuzzers achieve only 32\%, 60\%, and 65\%, respectively.

 \vspace{-0.3em}
\subsection{Bug Detection Performance (Answering RQ2)}
\label{performance_bug_detection}
We now evaluate the bug-finding ability of the studied tools on \textbf{D2}. We benchmark MuFuzz against state-of-the-art detectors, including static analyzers and fuzzers. We report the results of each tool in terms of true positives and false negatives. {Moreover, we discuss the false positive rate of MuFuzz.} We also count the number of timeout and error cases reported by each tool.

\textbf{True Positive Analysis.}\quad We manually identify the true vulnerabilities detected by each tool based on the labels in \textbf{D2}. Table~\ref{results} summarizes the experimental results, from which we can see that MuFuzz reports more true positives than other tools for all bug classes. Specifically, MuFuzz generates 27 true positives for \textbf{UE}, which is significantly better than other tools. In contrast, Mythril and Slither report only 10 and 2 true positives, while {sFuzz} and {ConFuzzius} generate 12 and 16 true positives, respectively. This is mainly due to the fact that conventional tools ignore handling exceptions for the return value of external calls. For \textbf{RE} and \textbf{SE}, MuFuzz reports 16 and 19 true positives, which is 4 and 8 more than the state-of-the-art tool IR-Fuzz. Notably, MuFuzz achieves higher coverage than existing fuzzers such as {sFuzz} and IR-Fuzz, and thus discovers more bugs that other fuzzers may not identify. Overall, MuFuzz detects 195 vulnerabilities, which is roughly 2.5$\times$ and 1.4$\times$ more than Mythril and IR-Fuzz, both of which are the best in their categories.

\textbf{False Negative Analysis.}\quad Further, we analyze the false negatives of each tool. It is clear from Table~\ref{results} that MuFuzz generates fewer false negatives compared to its counterparts. For example, MuFuzz produces only 3 false negatives for \textbf{IO}, while static tools such as Mythril and Osiris report 12 and 20 false negatives, respectively. This disparity can be attributed to the fact that: 1) static tools suffer from the inherent limitations of addressing complex program paths, which may encounter the path explosion problem, 2) they have difficulty detecting vulnerabilities that require to be triggered by dynamic execution, and 3) some of the bug oracles they rely on are unsatisfactory and imprecise, resulting in many false negatives.
Meanwhile, the fuzzers {sFuzz} and {Smartian} also generate 26 and 22 false negatives for \textbf{IO}, respectively, which may be due to their difficulties in fuzzing large contracts. Promisingly, MuFuzz reports only 20 false negatives and has no false negatives for five bug classes (\textbf{UD}, \textbf{RE}, \textbf{US}, \textbf{SE}, and \textbf{TO}), demonstrating its superiority in smart contract fuzzing.

\textbf{False Positive Analysis.}\quad The false positive rate of MuFuzz is extremely low and can be confidently disregarded. This can be attributed to the following reasons: 1) MuFuzz executes smart contracts in a true EVM virtual machine environment, which enables it to accurately capture issues in the execution of smart contract code, and 2) MuFuzz possesses the ability to adapt based on the actual execution of smart contracts, thereby enhancing its capacity to differentiate between normal behavior and abnormal scenarios triggered by vulnerabilities, which significantly contributes to the reduction of false positives. It is worth noting, however, that while MuFuzz maintains a low false positive rate, there may still be instances of potential false positives. We have conducted a manual audit and found that some false positives can be ascribed to the imprecise definition of some bug oracles, which occasionally result in false positives when analyzing the fuzzing report.

We now show the timeout and error cases reported by each tool. Here, {\small\ttfamily timeout} means that the analysis failed to converge within a 10-minute time frame, while {\small\ttfamily error} indicates that the analysis was aborted due to infrastructure issues, e.g., unsupported version or system internal bug. For example, Oyente and Osiris only support up to the version of {\small\ttfamily solc-0.4.19} and thus report error cases. Mythril reports 72 timeout cases due to the difficulty of traversing many execution paths in large contracts. In summary, our evaluation shows that MuFuzz can effectively detect various bugs in smart contracts, outperforming existing static analyzers and fuzzers.

\begin{figure}
    \centering
    \includegraphics[width=8.6cm]{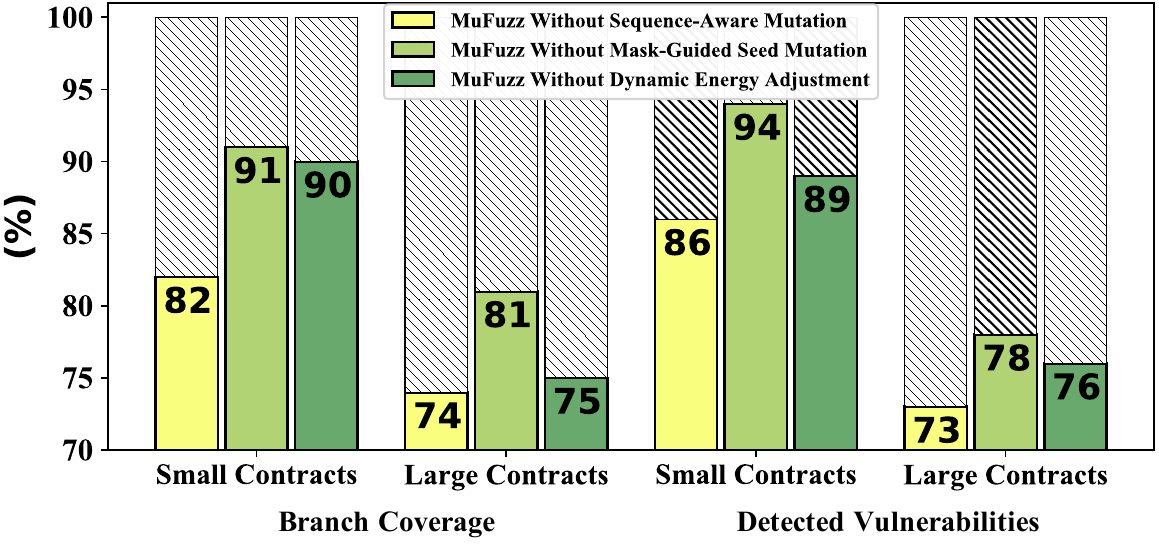}
    \vspace{-0.9em}
    \caption{Ablation study of each component in MuFuzz. }
     \label{component_results}
     \vspace{-1.8em}
\end{figure}

\subsection{Component Evaluation (Answering RQ3)}
\label{component_evaluation}
In the following, we further evaluate the importance of the proposed components in MuFuzz, including {sequence-aware mutation}, {mask-guided seed mutation}, and {dynamic energy adjustment}. Following the settings of the previous work~\cite{torres2021confuzzius}, we randomly select 100 contracts from \textbf{D1-small} and \textbf{D1-large}, respectively. Thereafter, we perform three experiments on each contract, where we deactivate a distinct component of MuFuzz for each experiment. 

First, we disable the sequence-aware mutation of MuFuzz (\S\ref{sequence_mutation}), but keep the other two components enabled. This means that MuFuzz generates transaction sequences with a random construction strategy, which we term as ``MuFuzz without sequence-aware mutation". We then modify MuFuzz by removing the mask-guided seed mutation (\S\ref{mask_mutation}), while using a random test input mutation. This variant is denoted as ``MuFuzz without mask-guided seed mutation". Finally, we replace the dynamic-adaptive energy adjustment (\S\ref{target_branch_revisting}) with a default allocation scheme used in {sFuzz}. This variant is referred to as ``MuFuzz without a dynamic energy adjustment".

Figure~\ref{component_results} shows the experimental results, where each bar represents the percentage of achieved coverage or detected vulnerabilities compared to the results when all three components were enabled (i.e., the gray bar in the back). It can be seen from the figure that each component is essential to the performance improvement of MuFuzz. In particular, we can observe that generating a meaningful transaction sequence plays the most critical role in achieving branch coverage and detecting more vulnerabilities. Quantitatively, without the three components, the achieved coverage decreases by 18\%, 9\%, 10\% on small contracts and 26\%, 19\%, 25\% on large contracts, respectively. Moreover, MuFuzz can discover 14\%, 6\%, 11\% more bugs on small contracts and 27\%, 22\%, 24\% more bugs on large contracts than without the three components, respectively.

\renewcommand\arraystretch{1.0}
\begin{table}
    \small
    \centering
    \caption{Real-World Case Studies of MuFuzz.} 
    \vspace{-0.9em}
    \begin{tabular}{ccll}
   \hline
     \textbf{Bug ID}   & \textbf{Reported Bugs}    &  \textbf{TP}    & \textbf{FP}  \\
    \hline
     \textbf{BD}   & 21    &  20   &  1 \\
     \textbf{UD}   & 0    &   0  &  0 \\
     \textbf{EF}   & 0    & 0    & 0 \\
     \textbf{IO}   &  42   &  42   & 0 \\
     \textbf{RE}   & 10    &   7  &  3 \\
     \textbf{US}   & 1    &   1  & 0 \\
     \textbf{SE}   & 2    &  2   & 0  \\
     \textbf{TO}   &  0   &  0   &  0  \\
     \textbf{UE}   & 10    &  9   & 1 \\
    \hline
    \textbf{Total} &  86   &  81  & 5  \\
    \hline
    \textbf{Average Coverage} &  \multicolumn{3}{c}{\textbf{80.71\%}}   \\
    \hline
    \end{tabular}
    \vspace{-2.2em}
        \label{case_study_results}
\end{table}

\subsection{Real-World Case Study (Answering RQ4)}
\label{case_study}
We also pay attention to the scalability and practicality of MuFuzz. We randomly select 100 contracts from \textbf{D3}, where each contract contains more than 30,000 transactions in Ethereum. We manually check the bug detection results and classify them into true positives (TP) and false positives (FP). In addition, we present the overall branch coverage (i.e., the average of the 100 contract runs) of MuFuzz. Table~\ref{case_study_results} summarizes the experimental results. From the table, we can see that: (1) MuFuzz reports a total of 86 bug alarms. Out of the 100 contracts, 39 contracts are flagged as having at least one of these alarms. We manually verify the alarms and confirm that 94\% of them are true positives. (2) MuFuzz achieves an average coverage of 80.71\% on the 100 contracts, showing inspiring practicality in testing real-world large contracts.

\textbf{Case Study.}\quad
To provide further insight on how MuFuzz achieves high coverage and detects bugs, we illustrate the fuzzing strategy of MuFuzz by through its workflow. Fig.~\ref{code:example} shows a contract where MuFuzz achieves 100\% coverage while sFuzz and ConFuzzius achieve only 50\%. Function {\small\ttfamily withdraw} has an if condition at line 30. If the condition cannot be satisfied, the bug inside the condition cannot be exposed. To cover line 31, one must call function {\small\ttfamily invest} twice to set the {\small\ttfamily phase} = 1. sFuzz and ConFuzzius fail to cover line 31 because they cannot generate a sequence that runs {\small\ttfamily invest} twice. MuFuzz, instead, incorporates a sequence-aware mutation strategy to create a sequence that can handle such conditions. Particularly, MuFuzz works in four main steps: 1) MuFuzz parses a contract source code into bytecode, ABI, and AST. By analyzing the ABI and AST, MuFuzz identifies which state variables are defined. 2) MuFuzz derives the dataflow dependencies of state variables and formulates a sequence of transactions. 3) MuFuzz determines the test inputs for each transaction in the sequence and activates the mutation masking strategy to evolve the seed inputs. 4) MuFuzz analyzes the fuzzing log and identifies if there is a vulnerability by matching it against the bug oracles. The vulnerability in the \emph{crowdsale} case can be exposed by MuFuzz with a sequence of transactions:
\begin{itemize}[topsep=1pt, itemsep=1pt]
\item $t_1$: A user calls {\small\ttfamily invest()} with {\small\ttfamily donations $\geq$ goal};
\item $t_2$: An attacker calls {\small\ttfamily invest()} again with a goal of setting {\small\ttfamily phase = 1};
\item $t_3$: An attacker calls {\small\ttfamily withdraw()}.
\end{itemize}

\section{Discussion}
\label{discussion}
We discuss the limitations and potential improvements of MuFuzz, and the significance of our work in data engineering.

\textbf{Data Dependency Analysis.}\quad MuFuzz approximates a transaction sequence by capturing the data dependencies of all state variables in a contract. However, the data flow facts collected from the data dependency analysis may contain spurious data flows that cannot occur in practice, which may degrade the fuzzing performance. In addition, MuFuzz may have difficulty in coping with complex and intertwined dependencies in the case of state variables across multiple functions. In the future, we will improve the data dependency analysis, making it work more effectively with fuzzing.

\textbf{Sequence Generation Analysis.}\quad In order to handle the \emph{persistent} state nature of smart contracts, generating a specific sequence of function calls is critical in smart contract fuzzing. For example, some existing works~\cite{nguyen2020sfuzz,choi2021smartian} start from a fresh state for each fuzz round with a transaction sequence as input. During the mutation phase, new transaction sequences are generated by mutating parts of the sequence. Such methods tend to incur high overhead in re-executing transactions to return back to a previous state. Instead, MuFuzz uses a sequence-aware strategy to generate and mutate the transaction sequence once, reducing the runtime overhead. However, MuFuzz may fail to explore some possible states due to the limited number of generated sequences. To this end, an interesting and promising future improvement is not to re-execute the previous transactions, but to move directly to some intermediate state and use it only to trigger new states.

\section{Related Work}
\label{related_work}

\subsection{Smart Contract Security Analysis} 
The rate of security incidents in smart contracts has sparked a strong interest in bug-finding in the community. Existing methods can be broadly classified into three categories.

\textbf{Static Analysis.}\quad In practice, static analysis can be broken down into program analysis, symbolic execution, formal verification, and machine learning. For example, Ethainter~\cite{brent2020ethainter} performs taint analysis to detect information flow bugs in smart contracts. Securify~\cite{tsankov2018securify} infers the semantic information of smart contracts by analyzing control- and data- dependencies, thus proving the presence or absence of bugs. To facilitate the analysis of smart contracts, a number of tools has tried customized intermediate representations, such as the XML parsing tree used in SmartCheck~\cite{tikhomirov2018smartcheck}, RBR used in {EthIR}~\cite{albert2018ethir}, XCFG used in ClairvOyance~\cite{ye2020clairvoyance}, and SSA used in Slither~\cite{feist2019slither}. Symbolic execution, as one of the preferred techniques for program analysis, has been widely used to detect smart contract bugs. Oyente~\cite{luu2016making} is one of the pioneering symbolic execution engines for smart contracts, followed by Osiris~\cite{torres2018osiris}, Manticore~\cite{mossberg2019manticore}, Mythril~\cite{mythril}, HoneyBadger~\cite{torres2019art}, DefectChecker~\cite{chen2021defectchecker}, and so on. Formal verification is also commonly used to validate the logical integrity of smart contracts, \emph{e.g.,} {VeriSmart}~\cite{so2020verismart} and {Zeus}~\cite{kalra2018zeus}. Recent years have witnessed an increasing practice of using machine/deep learning-based methods to detect smart contract vulnerabilities. Wesley et al.~\cite{tann2018towards} introduce a sequential learning approach for smart contracts using LSTM networks as a detection model. Contractward~\cite{wang2020contractward} extracts bigram features from opcodes and applies various machine learning algorithms to detect flaws in smart contracts. Liu et al.~\cite{liu2023combining} propose to translate the contract source code into a contract graph and build a graph model to identify vulnerabilities. 

\textbf{Dynamic Analysis.}\quad Dynamic analysis finds potential security problems in smart contracts by directly executing the contract code, typically via fuzzing test. {ContractFuzzer}~\cite{jiang2018contractfuzzer} is one of the earliest fuzzing frameworks for smart contracts, which identifies vulnerabilities by monitoring runtime behavior during a fuzzing campaign. Follow-up researchers further propose to study improvements to different parts of smart contract fuzzing. Recent fuzzers {ContraMaster}~\cite{wang2020oracle}, {Harvey}~\cite{wustholz2020harvey}, {Echidna}~\cite{grieco2020echidna} adopt a feedback-driven mechanism to generate diverse inputs that are more likely to reveal bugs. {sFuzz}~\cite{nguyen2020sfuzz} presents the idea of branch distance feedback to guide the fuzzer to explore hard-to-reach branches. Liu et al.~\cite{10018241} propose a sequence generation strategy that contains invocation ordering and prolongation to trigger deeper states.

\textbf{Hybrid Analysis.}\quad Fuzzing has evolved into a synthesis technique that synergistically combines static and dynamic analysis to generate better test cases and detect more bugs.
He et al.~\cite{he2019learning} present ILF, an end-to-end system that learns an effective and fast fuzzer from symbolic execution by phrasing the learning task in the framework of imitation learning. Xue et al.~\cite{xue2022xfuzz} propose a machine learning-guided smart contract fuzzing framework that uses machine learning predictions to guide fuzzers for vulnerability detection. {ConFuzzius}~\cite{torres2021confuzzius} presents a hybrid smart contract fuzzer, which adopts symbolic execution to search for uncovered branch conditions and creates meaningful transaction sequences of inputs at runtime by exploiting data dependencies. {Smartian}~\cite{choi2021smartian} utilizes data-flow-based feedback to effectively find meaningful transaction sequence orders. RLF~\cite{su2022effectively} introduces a reinforcement learning-guided fuzzing framework for smart contracts, which helps generate vulnerable transaction sequences.
MuFuzz outperforms such fuzzing tools by employing a sequence-aware mutation scheme to discover potential execution paths that cannot be identified by current smart contract fuzzers. In addition, the mask-guided seed mutation technique embedded in MuFuzz helps explore states in deeply nested branches.

\subsection{Fuzzing in Program Testing}
Fuzzing, which was initially put forth by Barton Miller~\cite{miller1990empirical}, becomes one of the most popular program testing techniques. A fuzzing campaign begins by generating a number of test inputs for the target programs, and tries to uncover bugs by monitoring the execution states of the program during runtime. Depending on how much information is required from the target program at runtime, fuzzing techniques can be divided into three categories: \emph{black-box, white-box}, and \emph{grey-box}~\cite{liang2018fuzzing}. 

Black-box fuzzing, which does not rely on any information from the target program, employs predefined rules to randomly mutate a given seed for generating test inputs. Black-box fuzzers are popular in program analysis for their effectiveness in bug discovery and ease of use. Trinity~\cite{trinity} is a black-box fuzzer used to fuzz system call interfaces in Linux kernels. DiffChaser~\cite{xie2019diffchaser} introduces a black-box testing framework to identify discrepancies between different versions of DNN. FuzzSMT~\cite{brummayer2009fuzzing} and StringFuzz~\cite{blotsky2018stringfuzz} use grammar-based black-box fuzzing to create valid SMT formulas.

Instead, white-box fuzzing relies on the internal logic within the target programs. It progressively examines statements, conditions, execution paths, data flow, and a wide range of normal and abnormal inputs to assess if the program yields expected results. One of the pioneering white-box fuzzers is SAGE~\cite{godefroid2008automated}, which incorporates multiple optimizations to handle numerous execution paths in Windows applications. BuzzFuzz~\cite{ganesh2009taint} introduces a taint-based white-box fuzzing technique to automatically generate test inputs that exercise code deep within the semantic core of programs. Recently, DeepXplore~\cite{pei2017deepxplore} proposes a white-box testing framework to identify behavioral inconsistencies between different DNNs.

Grey-box fuzzing occupies a middle ground between black-box and white-box, and has gained popularity due to its strong performance and reasonable runtime overhead. Grey-box fuzzing falls into two kinds: coverage-guided greybox fuzzing (CGF) and directed greybox fuzzing (DGF)~\cite{zong2020fuzzguard}. AFL~\cite{afl} serves as a prominent example of CGF fuzzers. It adopts a lightweight instrumentation technique and a genetic algorithm to improve coverage. AFLGO~\cite{bohme2017directed} represents a cutting-edge DGF fuzzer. It computes the distance between entry points, guiding the seed mutation to cover target locations in the control flow graph where faulty code resides.

\section{Conclusion}
\label{conclusion}
In this paper, we present MuFuzz, a novel fuzzing framework for smart contracts. To increase the probability of exploring deeper states, we introduce a sequence-aware mutation and seed mask guidance strategy, enforcing the fuzzer to trigger complex states guarded by strict branch conditions. To make the fuzzing resources for each branch more balanced, we develop a dynamic energy adjustment mechanism that can adjust the energy allocation according to the weight value of each branch during fuzzing. It is worth noting that the techniques proposed in MuFuzz can also be transferable to the fuzzing of other programs. Extensive experiments on three benchmarks show that our system performs better than state-of-the-art smart contract analyzers in terms of coverage, efficiency, and bug-finding performance. We publish both our system and benchmarks, hoping to push forward future research.

 \vspace{-0.2em}
\section*{Acknowledgment}
 \vspace{-0.1em}
This work was supported by the National Key R\&D Program of China (2021YFB2700500), the Key R\&D Program of Zhejiang Province (2022C01086).

\bibliographystyle{IEEEtran}
\bibliography{conference_101719}

\end{sloppypar}
\end{document}